\documentclass[useAMS,usenatbib,usegraphicx]{mn2e}
\usepackage{color,amsmath}
\DeclareGraphicsExtensions{.pdf,.png,.jpg,.ps}
\usepackage{epsfig}

\newcommand{\NIRdist}[1][24.51 $\pm$ 0.08]{#1}  
\newcommand{\VISdist}[1][24.32 $\pm$ 0.09]{#1}  

\begin{document}

\title{Panchromatic Hubble Andromeda Treasury XIII: The Cepheid Period-Luminosity Relation in M31}

\author[Wagner-Kaiser et al.]{R.~Wagner-Kaiser,$^{1}$ A.~Sarajedini,$^{1}$ J. J.~Dalcanton,$^{2}$ B. F.~Williams,$^{2}$ A.~Dolphin,$^{3}$\\
$^1$University of Florida, Department of Astronomy, 211 Bryant Space Science Center, Gainesville, FL, 32611\\
$^2$University of Washington, Department of Astronomy, 3910 15th Ave NE, Seattle, WA, 98195\\
$^3$Raytheon Company, 1151 E Hermans Rd, Tucson, AZ 85756\\
}

\date{}

\pagerange{\pageref{firstpage}--\pageref{lastpage}} \pubyear{2014}

\maketitle

\label{firstpage}


\begin{abstract}
Using Hubble Space Telescope Advanced Camera for Surveys (HST/ACS) and Wide Field Camera 3 (WFC3) observations from the Panchromatic Hubble Andromeda Treasury (PHAT), we present new period-luminosity relations for Cepheid variables in M31. Cepheid from several ground-based studies are identified in the PHAT photometry to derive new Period-Luminosity and Wesenheit Period-Luminosity relations in the NIR and visual filters. We derive a distance modulus to M31 of \NIRdist{} in the IR bands and \VISdist{} in the visual bands, including the first PL relations in the F475W and F814W filters for M31. Our derived visual and IR distance moduli disagree at slightly more than a 1-$\sigma$ level. Differences in the Period-Luminosity relations between ground-based and HST observations are investigated for a subset of Cepheids. We find a significant discrepancy between ground-based and HST Period-Luminosity relations with the same Cepheids, suggesting adverse effects from photometric contamination in ground-based Cepheid observations. Additionally, a statistically significant radial trend in the PL relation is found which does not appear to be explained by metallicity. 
\end{abstract}

\vspace{2pc}


\section{Introduction}\label{Intro}

The pulsating class of variable stars known as Cepheids has been studied for a number of reasons, the most prominent of which is the existence of a relation between their periods and luminosities. The Period-Luminosity (PL) relation for Cepheid variables casts these stars as standard candles, making them ideal for determining distances. As a result, Cepheids play a vital role in the cosmological distance ladder, being observable from the Local Group to distances of tens of Mpc, and overlapping with secondary distance indicators. 

More accurate distances from the Cepheid PL Relation lead to improved calibration of stellar luminosities, constraints on stellar population synthesis models, and measurements of the Hubble Constant (H$_{0}$). Improved precision of the Period-Luminosity relation has become increasingly important in light of the tension between the Cepheid-based determination of H$_{0}$ and that from the Planck mission (Riess et al. 2011, Planck Collaboration 2013).

The possibility of using M31 as a local anchor for the Period-Luminosity relation in the future could benefit the effort to improve the distance scale. Although the further distance degrades accuracy of measurements compared to more local distance estimates ($\it{i.e.}$, Milky Way and Large Magellanic Cloud), M31 has the opportunity to diversify the methods used to determine distances. Compared to the Large Magellanic Cloud, M31 is more akin to external galaxies used for even further distance measurements. The use of M31 as an anchor could potentially benefit distance determinations for the extragalactic distance scale, and perhaps assist in addressing the discrepancy between H${_0}$ values.

The Panchromatic Hubble Andromeda Treasury (PHAT) survey of M31 provides a unique opportunity to image a significant number of Cepheids previously only observed from ground-based telescopes. Obtaining accurate Hubble Space Telescope (HST) magnitudes for these variable stars promises to reduce the error in the PL relation for M31, leading the way to a more precise value of the distance. Previous ground-based surveys have detected thousands of Cepheids in and near the M31 disk ($\it{e.g.}$, Stanek et al. 1998; An et al. 2004; Kodric et al. 2013), but prior to PHAT, only small samples of Cepheids in M31 have been observed with HST (Macri et al. 2001; Riess et al. 2012). The extensive coverage of the PHAT program opens new opportunities to improve the distance to M31 via observations of larger numbers of Cepheid variables with HST.

There are three main contributors to the Cepheid distance uncertainty: metallicity effects, blending and crowding, and the uncertainty in the distance to the LMC (Riess et al. 2009; Freedman \& Madore 2010a). The effort to reduce the error and dispersion in the PL relation has made large gains in the past two decades (Madore \& Freedman 1992, Macri 2005, Freedman \& Madore 2010), though further improvements are on the horizon (Gerke et al. 2011).

There is as yet no clear consensus on the universality of the PL relation and its dependence (Bono et al. 2008), or lack thereof (Majaess 2011), on different passbands or on the metallicity of the stellar population. If metallicity does affect the PL relation, any uncertainties are likely to be minimized in the near-infrared (NIR) bands as compared to the visual (Madore \& Freedman 1991). Additionally, obtaining Cepheid PL relations in the NIR reduces the impact of dust on magnitude measurements. Both metallicity effects and extinction effects have driven the push towards observations of Cepheids at longer wavelengths.

Blending and crowding can contaminate the photometry of Cepheids, making true magnitudes more difficult to obtain. This problem is significant for ground-based observations, which may be biased by up to 0.2 magnitudes by blending (Mochejska et al. 2000, Vilardell et al. 2007). However, point-spread function (PSF) magnitudes obtained from HST greatly reduce these effects, thereby reducing the error in the measurements of Cepheid magnitudes and thus the uncertainty in the distance and Hubble constant as well.  Furthermore, making use of a Wesenheit magnitude (an index combining magnitude and color) can help to account for reddening variations from star to star by taking individual Cepheid colors into account (Madore 1982; Opolski 1983; Moffett \& Barnes 1986; Madore \& Freedman 1991; Caputo et al. 2000; Leonard et al. 2003; Ngeow \& Kanbur 2005; Fiorentino et al. 2007; Bono et al. 2008, 2010; Ngeow 2012).

Lastly, uncertainty in the LMC distance is thought to account for about 5\% percent of the uncertainty in the cosmological distance scale (Freedman \& Madore 2010a; Riess et al. 2011; Freedman et al. 2012; Inno et al. 2013). By using NGC 4258 as an anchor for the distance scale instead of the LMC, this uncertainty can be reduced to slightly more than 3\% (Riess et al. 2009b; Riess et al. 2011; Macri et al. 2006; Fiorentino et al. 2013). Humphreys et al. (2013) have used masers to update the distance to NGC 4258 to 7.60$\pm$0.17$\pm$0.15 Mpc, which reduces the error of using the LMC as an anchor as well as reducing the error to NGC 4258 from previous studies (Herrnstein 1999).

In this work, we present magnitudes of Cepheids in M31 from the PHAT survey and use PL relations to redetermine the distance modulus of M31. Section \ref{Data} presents the details of the data we have analyzed. The methods and the analysis of the data to construct PL relations are discussed in Section \ref{Results}, where we decrease the dispersion in the visual PL relations and increase the sample of Cepheids with Near-Infrared (NIR) photometry with respect to Riess et al. (2012) and Kodric et al. (2015). In Section \ref{DIRECTPL}, we examine the benefits from the photometric precision from the PHAT survey. As we show by examining a subset of our Cepheids which were also observed by the DIRECT survey (see Section \ref{Data} for details), magnitudes obtained via HST in the visual bands are superior to recent ground-based surveys (e.g. Stanek et al. 1998; An et al. 2004; Kodric et al. 2013). In Section \ref{dists}, we discuss the determination of distances from the PHAT photometry. The subset of stars from DIRECT is used to compare PHAT determined distances to ground-based distance estimates. We explore the relationship between metallicity, distance, and radial location for the Cepheids in Section \ref{Distgrad}. The implications of our results are described in Section \ref{Discussion} while the conclusions are summarized in Section \ref{Conclusion}. 


\section{Data}\label{Data}

\begin{figure}
\centering
\includegraphics[scale=0.37]{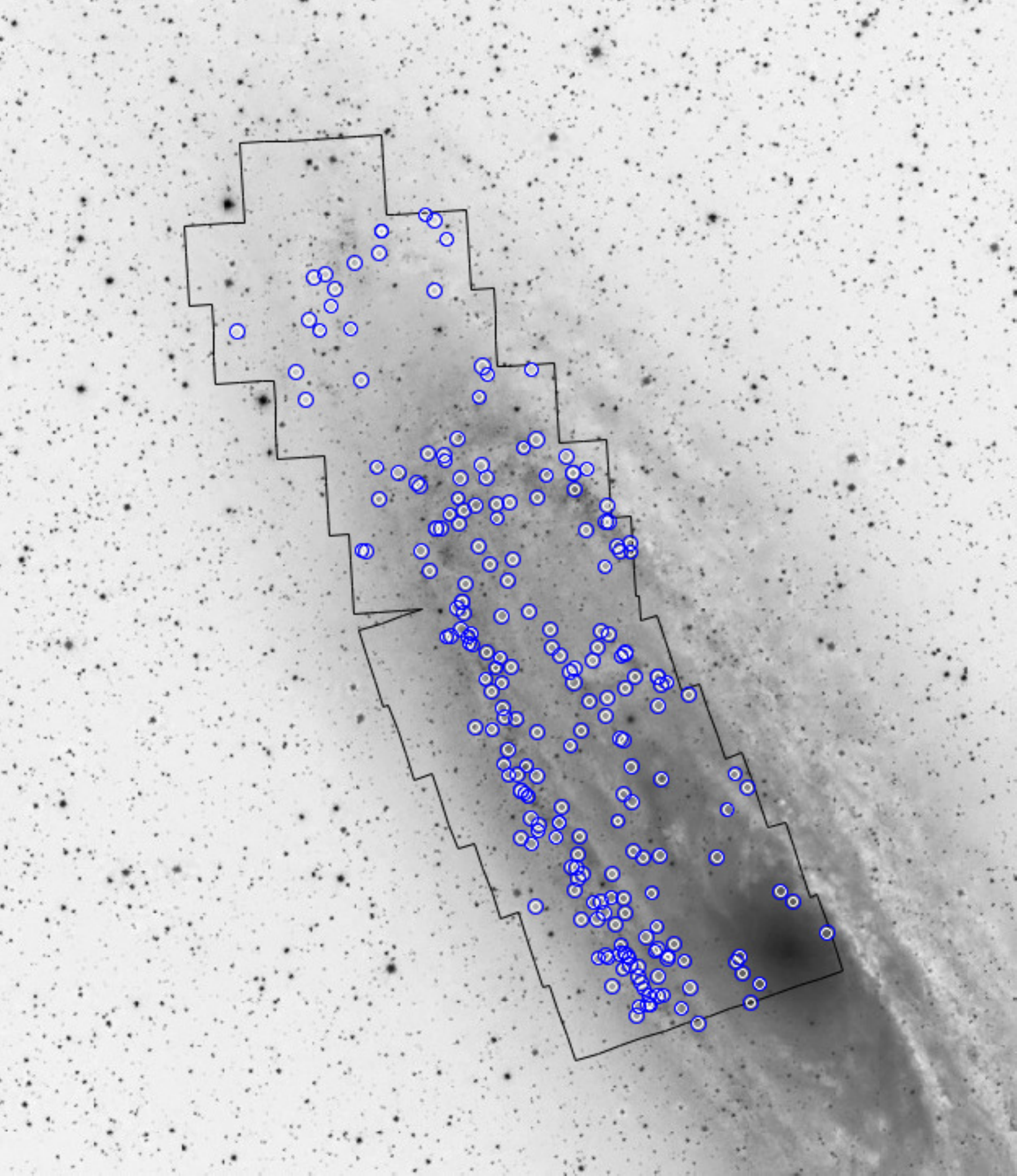}
\caption{The locations of 175 Cepheids from the complete dataset are shown as circles and the solid line shows the outline of the PHAT footprint and coverage of M31.}
\label{locations}
\end{figure}

The data used in this paper are obtained from the PHAT photometry of M31. The PHAT survey coverage, design, and photometry are described at length in Dalcanton et al. (2012). The observations utilize the Wide Field Camera 3 (WFC3) in both the ultraviolet-visible (UVIS) mode and the infrared (IR) mode and the Advanced Camera for Surveys (ACS) Wide Field Channel (WFC). They cover about one third of the disk of M31 across 6 filters (F275W and F336W in the UV with WFC3; F110W and F160W in the IR with WFC3; F475W and F814W in the visual bands with ACS). Each HST pointing is two orbits: 1 orbit for WFC3/UVIS and 1 orbit for WFC3/IR (with ACS/WFC in parallel mode). The coverage of M31 by PHAT is divided into 23 ``bricks'', each composed of 18 HST fields of view (in a 6$\times$3 layout, each brick covering a 3$\times$1.5 kpc area). Photometry is performed on each field and brick using DOLPHOT (Dolphin 2000) and the subsequent photometry files are filtered to reject low-quality and non-stellar objects, as described in Dalcanton et al. (2012) and Williams et al. (2014).

To obtain the largest sample of Cepheid observations possible, we use multiple Cepheid catalogs: PAndromeda (Kodric et al. 2013), DIRECT (Kaluzny et al. 1998, Stanek et al. 1998, Kaluzny et al. 1999, Stanek et al. 1999, Bonanos et al. 2003), and Cepheids already identified in the PHAT images (Riess et al. 2012).

The DIRECT survey imaged M31 from 1996 to 1997 with the McGraw Hill Telescope at the Michigan-Dartmouth-MIT Observatory (MDM) and from 1996 to 2000 with the 1.2 m telescope at the F. L. Whipple Observatory (FLWO). The Cepheids in the DIRECT survey were observed more than 130 times in the Johnson V-band but less frequently in the Johnson B and Cousins-Kron I filters. All stars have V photometry, but not necessarily B or I magnitudes. 6 fields were imaged by the survey, 5 of which overlap with the PHAT coverage of M31, including 87 Cepheids. Further details about the acquisition and reduction procedures can be found in the DIRECT papers (Kaluzny et al. 1998, Stanek et al. 1998, Kaluzny et al. 1999, Stanek et al. 1999, Bonanos et al. 2003).

The Pan-STARRS survey of M31, also referred to as PAndromeda, uses the 1.8 meter Panoramic Survey Telescope and Rapid Response System with the Giga Pixel Camera in Haleakala, Maui, Hawaii (Hodapp et al. 2004, Kaiser et al. 2002, Tonry \& Onaka 2009). From mid-2010 to late-2011, 183 epochs of data were gathered with a half hour of observing each night in the r$_{P1}$ and i$_{P1}$ (where P1 refers to the filter set used on the Pan-STARRS 1 system). Details of the observations as well as reduction procedures can be found in Kodric et al. (2013) and Lee et al. (2012). A similar study examining these stars in the NIR filters from the PHAT images from MAST can be found in Kodric et al. (2015), though with a smaller sample size of about 111 long-period Cepheids.

Photometry of 67 Cepheids from the first year of PHAT data were examined by Riess et al. (2012). The initial identification of these Cepheids was through the POMME Survey (Fliri et al. 2012), with an additional Cepheid from the DIRECT Survey (Kaluzny et al. 1998, Stanek et al. 1998, Kaluzny et al. 1999, Stanek et al. 1999, Bonanos et al. 2003), and two more Cepheids from the PAndromeda Survey (Kodric et al. 2013).

\begin{figure}
\includegraphics[scale=0.45]{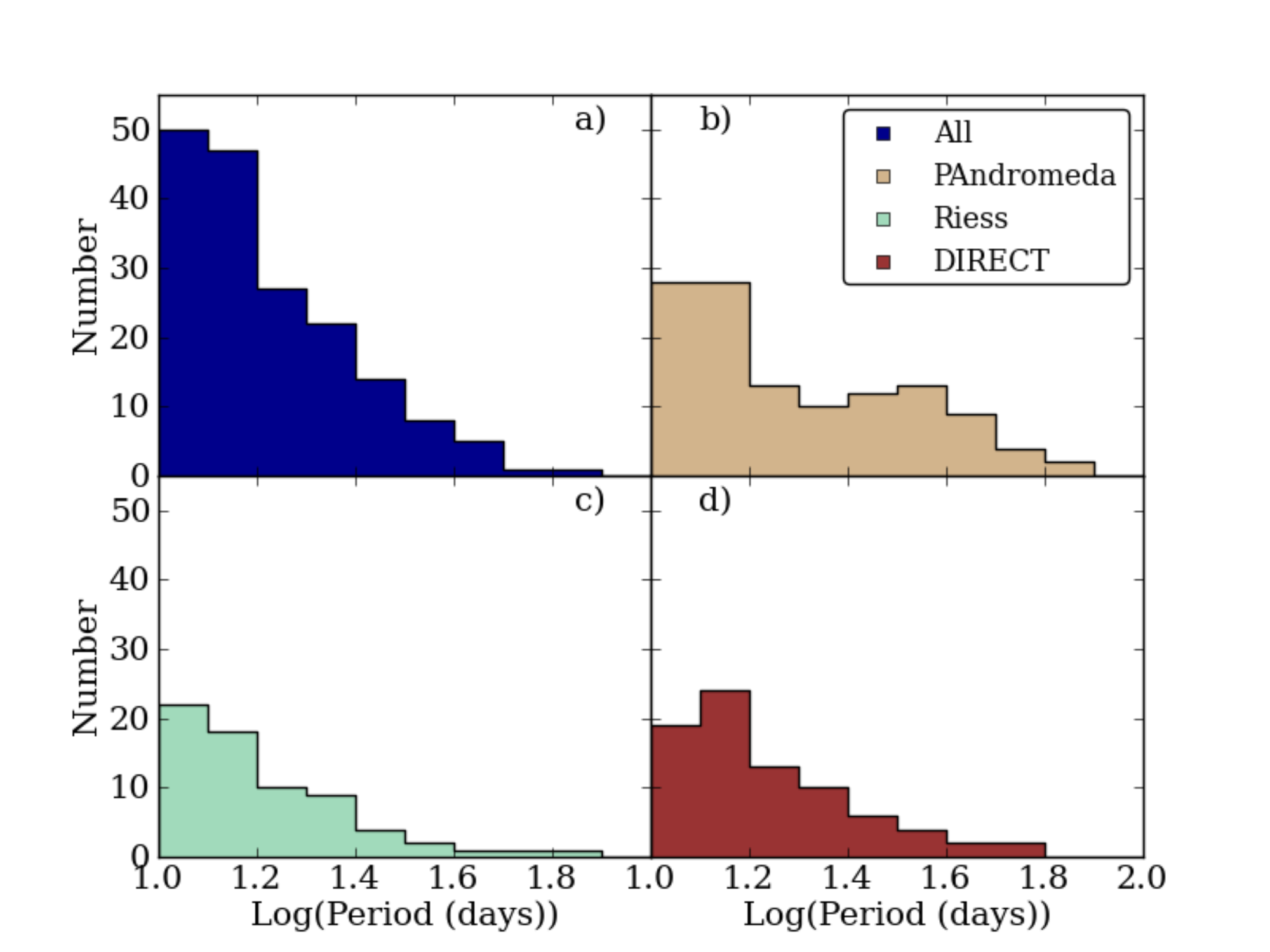}
\caption{The original period distributions of each (non $\sigma$-clipped) sample. Panel a) shows the complete dataset; panel b) gives the period distribution for PAndromeda Cepheids; panel c) shows the Riess et al. (2012) sample of periods; panel d) gives the period distribution of the DIRECT Cepheids. The periods are published values from the respective sources seen in Table \ref{Cepheidlist}.}
\label{periods}
\end{figure}

Using all three sources, we produce a dataset containing a total of 175 distinct variables with published periods longer than 10 days, detailed in Table \ref{Cepheidlist} and whose locations are shown in Figure \ref{locations}. We restrict our sample of Cepheids to those with periods longer than 10 days because the presence of a break in linearity in the P-L relation at a period of approximately 10 days has been well documented (Simon \& Lee 1981; Tammann \& Reindl 2002; Tammann et al. 2002; Kanbur \& Ngeow 2004; Sandage et al. 2004; Ngeow et al. 2005; Kodric et al. 2015). The distribution of these periods, separately by publication and altogether, is shown in Figure \ref{periods}.

To construct the dataset used in this paper, Cepheid locations were extracted from the PAndromeda and DIRECT surveys as well as from Riess et al. (2012). The positions were used to locate the photometry in the PHAT database. Because it is the largest dataset, positions were taken from the PAndromeda survey first, then from Riess et al. (2012), ignoring duplicates from PAndromeda, and lastly from the DIRECT survey, ignoring duplicates from the other two surveys. We adopt periods from these surveys in the same sequence.

\begin{figure}
\includegraphics[scale=0.45]{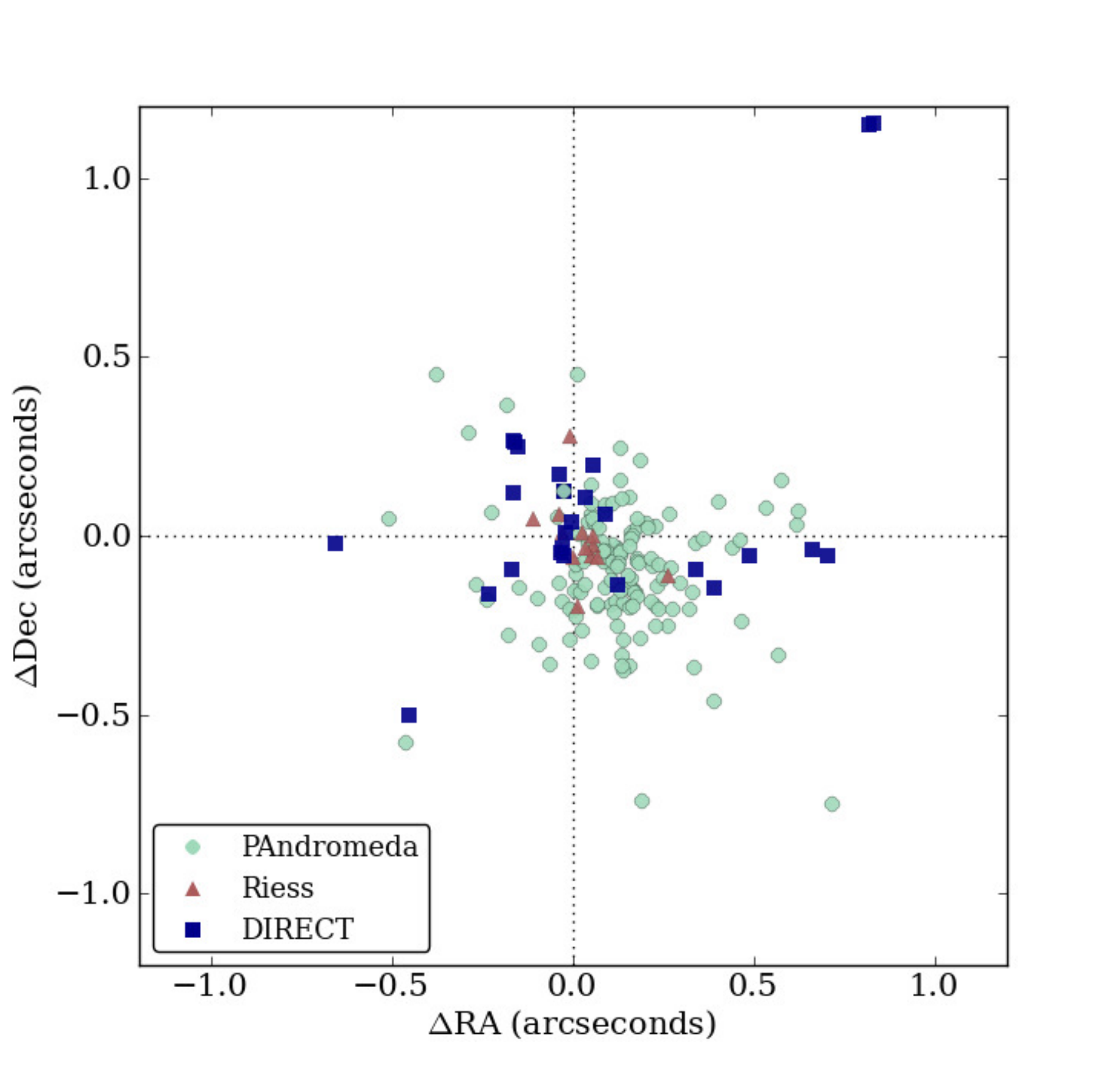}
\caption{A comparison of right ascension and declination in the PHAT photometry to the published values of the PAndromeda, Riess et al. (2012), and DIRECT surveys (i.e.: RA$_{PHAT}$-RA$_{survey}$). Green circles denote the comparison to PAndromeda, blue squares denote the comparison to the DIRECT survey, and the red triangles show the comparison to Riess et al. (2012). Several Cepheids only found in the DIRECT et al. dataset are outside the initial matching radius of 1";  no match was found until the matching radius was expanded to 1.2", allowing a unique match. We find a mean offset of 0.11'' in right ascension and --0.06'' in declination.}
\label{locationsdiff}
\end{figure}

Because the FWHM of the ground-based photometry is significantly larger than that of HST, there can be multiple sources that are plausible matches based solely on position. To identify the most likely Cepheid counterpart in the HST data, we first identify sources within 1" of the published right ascension and declination. Of these, the final match was based on choosing sources with magnitudes within $\pm$1 magnitude of the source catalog. Objects were rejected if they did not show variability beyond the photometric error in two PHAT epochs. Objects were also thrown out if there were multiple objects within 1 magnitude of the expected, published magnitude(s). For two objects published only in the DIRECT dataset, the matching radius was extended to 1.2" when no match was initially found within 1" (the expansion to a larger radii is not surprising given the large FWHM of the DIRECT photometry). A comparison of the PHAT Cepheids locations to their published locations is shown in Figure \ref{locationsdiff}. The mean offsets are approximate 0.11'' in right ascension and --0.06'' in declination.

Through this method, magnitudes in the PHAT dataset were obtained for each Cepheid in each available visual and NIR filter. The total sample is composed of 175 Cepheid variables with visual and 174 Cepheids with NIR magnitudes (see Table \ref{Cepheidlist}; one Cepheid is on the very edge of the PHAT coverage and falls outside the WFC3/IR footprint). The locations of the Cepheids as determined by PHAT are given in columns 1 and 2 of Table \ref{Cepheidlist}. In columns 3-10 of the table, we give the PHAT magnitudes in the visual filters (F475W and F814W) and the infrared filters (F110W and F160W), along with their photometric uncertainties from photo counts (uncertainties can be significantly larger due to crowding; see Dalcanton et al. 2012 ). In column 11, we give the published period of each Cepheid as drawn from by the reference listed in the final column.

The magnitudes obtained from the PHAT photometry are not time-averaged mean magnitudes, due to insufficient temporal coverage. These magnitudes are therefore snapshots of the Cepheid at a single phase in its variation. These ``random" phase magnitudes may lead to deviations from a given PL relation due to the amplitudes of the Cepheids' variation and the phase of observation.

The complete dataset is used for the primary analysis of this paper. However, as we describe in Section \ref{DIRECTPL}, we separately analyze the DIRECT sample to examine differences in the PL relations resulting from ground-based and space-based photometry. The properties of the DIRECT sub-sample is seen in Table \ref{DIRECTCephs}. The DIRECT variable name is given in column 1, with the right ascension and declination as determined by the PHAT data given in columns 2 and 3. The F475W and F814W magnitudes from the PHAT survey are given in columns 4 and 6 with their corresponding photometric errors (columns 5 and 7). Column 8 gives the period published by the DIRECT survey. Out of the 85 Cepheids in the DIRECT sample whose positions overlap with the PHAT footprints, only 80 are used due to matching problems, equivalent to a 6\% rejection rate.

\begin{table*}
\centering
\caption{Complete Cepheid Sample}
    \begin{tabular}{@{}llllllllllll@{}}
    \hline
 \textbf{RA} & \textbf{Dec} & \textbf{F475W}& \textbf{$\sigma_{F475W}$}	& \textbf{F814W}& \textbf{$\sigma_{F814W}$}	& \textbf{F110W}& \textbf{$\sigma_{F110W}$}	& \textbf{F160W}& \textbf{$\sigma_{F160W}$}	& \textbf{Period (days)}& \textbf{Source} \\  
\hline
10.91937 & 41.21261 & 22.948 & 0.052 & 20.682 & 0.040 & 19.982 & 0.073 & 19.303 & 0.100 & 10.045 & PAndromeda \\
11.27651 & 41.69481 & 20.764 & 0.037 & 19.608 & 0.053 & 19.333 & 0.057 & 18.883 & 0.046 & 10.054 & PAndromeda \\
11.01628 & 41.62777 & 21.076 & 0.009 & 19.796 & 0.033 & 19.213 & 0.081 & 18.786 & 0.088 & 10.101 & PAndromeda \\
11.46423 & 42.14078 & 21.448 & 0.006 & 19.823 & 0.004 & 19.173 & 0.052 & 18.719 & 0.063 & 10.234 & PAndromeda \\
11.02203 & 41.23451 & 22.039 & 0.036 & 20.254 & 0.025 & 19.686 & 0.105 & 19.134 & 0.111 & 10.29 & DIRECT \\
11.09179 & 41.35495 & 21.736 & 0.014 & 19.940 & 0.021 & 19.490 & 0.101 & 18.989 & 0.114 & 10.296 & PAndromeda \\
11.18490 & 41.92719 & 21.163 & 0.005 & 19.796 & 0.017 & 19.452 & 0.097 & 18.895 & 0.108 & 10.3 & Riess \\
11.17153 & 41.40722 & 22.208 & 0.037 & 20.239 & 0.024 & 19.399 & 0.100 & 18.858 & 0.105 & 10.35 & DIRECT \\
11.34817 & 42.03007 & 21.452 & 0.034 & 19.829 & 0.017 & 19.666 & 0.027 & 19.076 & 0.038 & 10.371 & PAndromeda \\
11.38078 & 41.88077 & 21.902 & 0.035 & 20.033 & 0.020 & 19.530 & 0.044 & 19.056 & 0.070 & 10.43 & Riess \\
\hline \\
    \end{tabular}
   \label{Cepheidlist}
\end{table*}

\begin{table*}
\centering
\caption{DIRECT Cepheid Sample}
    \begin{tabular}{@{}llllllll@{}}
    \hline
 \textbf{DIRECT ID}  &  \textbf{RA} & \textbf{Dec} & \textbf{F475W}& \textbf{$\sigma_{F475W}$}	& \textbf{F814W}& \textbf{$\sigma_{F814W}$}	& \textbf{Period (days)} \\  
\hline
V5343 & 11.02203 & 41.23451 & 22.039 & 0.036 & 20.254 & 0.025 & 10.29 \\
V8515 & 11.09179 & 41.35495 & 21.736 & 0.014 & 19.940 & 0.021 & 10.308 \\
V13153 & 11.17153 & 41.40722 & 22.208 & 0.037 & 20.239 & 0.024 & 10.35 \\
V2293 & 11.13846 & 41.62605 & 21.120 & 0.004 & 19.548 & 0.006 & 10.567 \\
V6363 & 11.36852 & 41.65947 & 22.249 & 0.008 & 20.258 & 0.007 & 10.593 \\
V410 & 11.09179 & 41.66419 & 21.800 & 0.035 & 19.999 & 0.020 & 10.792 \\
V13042 & 11.16980 & 41.39944 & 21.772 & 0.035 & 20.141 & 0.022 & 10.847 \\
V3773 & 10.98764 & 41.24036 & 23.130 & 0.070 & 19.874 & 0.020 & 10.938 \\
V7381 & 11.09263 & 41.32135 & 20.808 & 0.029 & 19.586 & 0.019 & 10.943 \\
V4733 & 11.33250 & 41.78882 & 22.337 & 0.006 & 20.241 & 0.006 & 10.971 \\
\hline \\
    \end{tabular}
    \label{DIRECTCephs}
\end{table*}

\subsection{Uncertainty Determination}\label{error}

We use artificial star tests to determine the true photometric uncertainties in all 4 relevant filters. For each field and camera in the PHAT survey, 10$^5$ artificial stars have been inserted and reprocessed through the PHAT pipeline (as detailed in Dalcanton et al. 2012 and Williams et al. 2014). Artificial stars are inserted individually and the photometry is re-run in the immediate vicinity of that star and the resulting magnitude is recorded. This procedure allow us to fully characterize non-trival noise from the photometric measurements, which in particular includes blends and completeness

From the artificial star tests for the corresponding brick and field for each Cepheid, we choose artificial stars with a recovered magnitude within 0.5 magnitudes of the Cepheid's observed magnitude and locations within 20 arcseconds of the Cepheid's location. The estimated dispersion and systematics of our measurements are then added in quadrature to produce the total photometric uncertainty for each Cepheid.

\subsection{Comparison with Riess et al. (2012)}\label{Riess2012comp}

\begin{figure}
\includegraphics[scale=0.45]{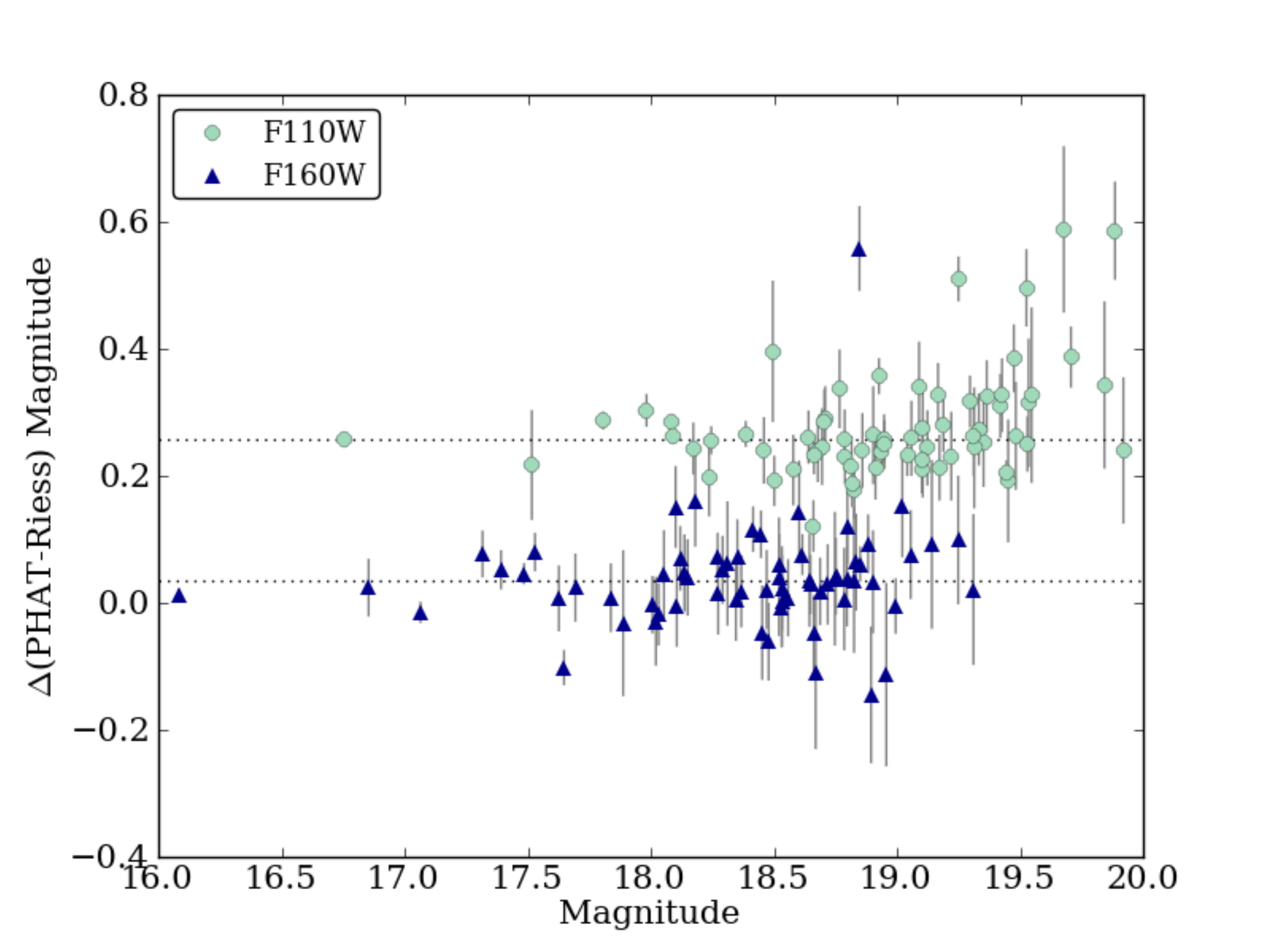}
\caption{A comparison of the magnitudes from PHAT photometry and Riess et al. (2012) photometry, with F110W denoted as green circles and F160W as blue triangles. The median difference in the F110W photometry is 0.259$\pm$0.028 magnitudes, a non-negligible offset. The median difference in F160W is only 0.035$\pm$0.010 magnitudes.}
\label{magcomp}
\end{figure}

There are 67 Cepheids in Riess et al. (2012) for which we have independently obtained photometry from the PHAT photometry pipeline. We compare the F110W and F160W magnitudes for each of these Cepheids in Figure \ref{magcomp}. We find a median offset of 0.259$\pm$0.028 in F110W and 0.035$\pm$0.010 in F160W. These offsets are similar to those noted by Kodric et al. (2015).  Our F160W photometry and that of Kodric et al. (2015) differ by an average of 0.016 and our F110W photometry differs by 0.001. Kodric et al. (2015) and the PHAT team both utilize PSF photometry and thus achieve very similar results. The discrepant F160W point seen in Figure \ref{magcomp} is the same Cepheid shown to be a misidentification in Kodric et al. (2015). 

\begin{figure}
\includegraphics[scale=0.45]{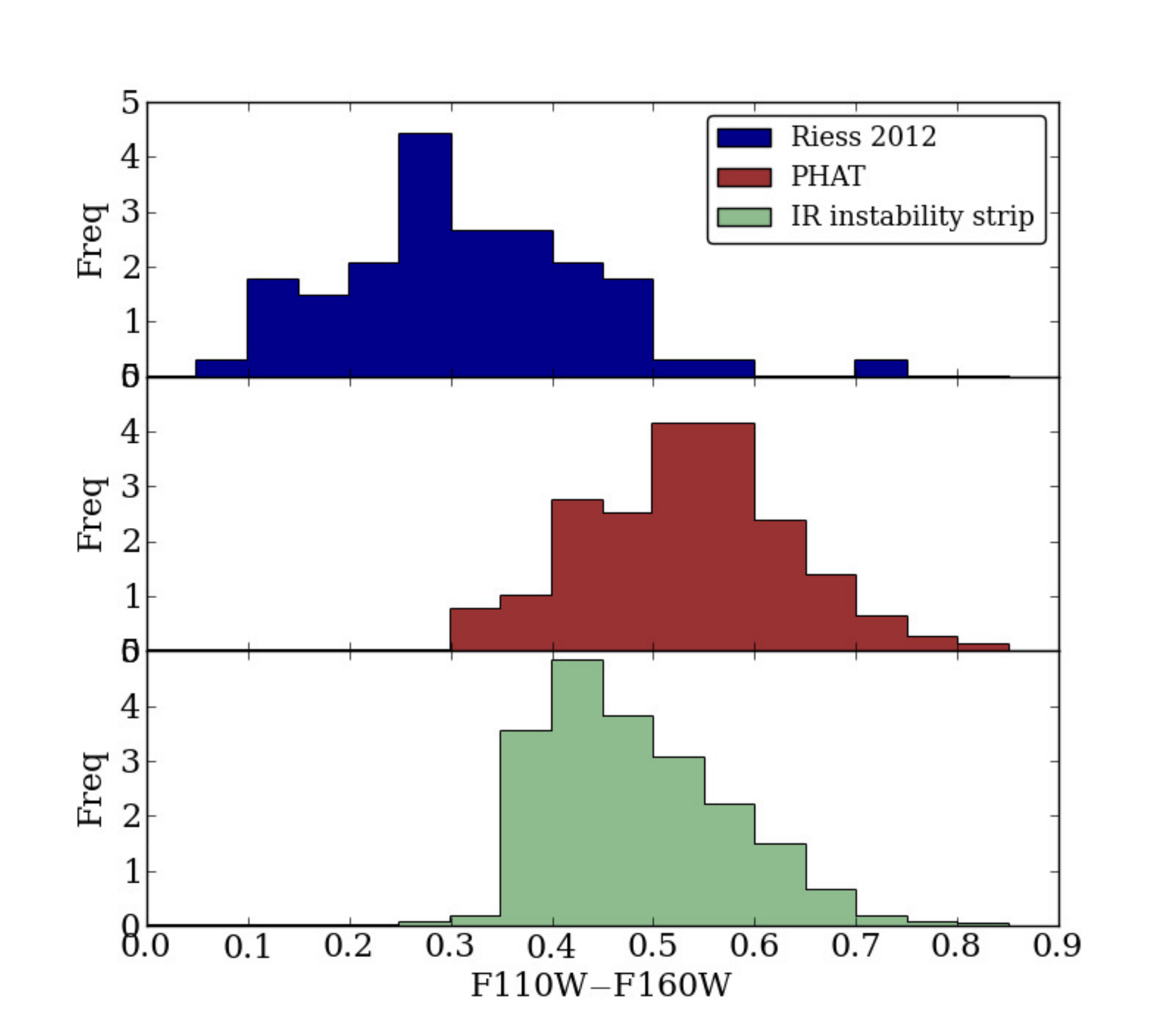}
\caption{We compare the F110W$-$F160W colors of Riess et al. (2012) (top panel), our PHAT photometry (middle panel), and theoretical colors (bottom panel). The theoretical colors are derived from a grid of Girardi isochrones over an age range of 4 Myr to 1 Gyr and metallicity range of Z=0.0001 to Z=0.05, constrained to the canonical instability strip defined in Bono et al. (2005). The mean of the Riess et al. IR colors is 0.32; for the PHAT colors the mean is 0.53; the mean derived from isochrones is 0.48. The theoretical F110W$-$F160W colors derived from the grid of isochrones is much more closely matched in mean and in range by our PHAT photometry than by the Riess et al. (2012) photometry.}
\label{IScomp}
\end{figure}

Figure \ref{IScomp} shows the F110W and F160W color differences for the Riess et al. (2012) photometry, the PHAT photometry, and theoretical colors from a grid of isochrones. Our set of models is composed of Girardi isochrones over an age range of 4 Myr to 1 Gyr and metallicity range of Z=0.0001 to Z=0.05, constrained to the canonical instability strip defined in Bono et al. (2005). The Riess et al. photometry is shown in the top panel and the PHAT photometry in the middle panel; the offset between the two observed datasets is clear. The offset is due to an ensquared energy fraction correction rather than an encircled energy fraction correction, the latter of which is the correct application for PSF photometry (Kodric et al. 2015). Whereas the ensquared energy fraction gives the energy over a certain number of pixels, the encircled energy fraction is the light within a certain radii. The difference between using the two energy fraction corrections is $\sim$0.258, which accounts for the majority of the discrepancies in photometry we see between our photometry and Riess et al. (2012). The range and mean of F110W$-$F160W colors observed in the PHAT photometry mimic those seen in the theoretical colors in the bottom panel of Figure \ref{IScomp}.

It is important to note this offset between Riess et al. (2012) and our photometry, as it informs the process by which we determine a distance with the NIR filters (Section \ref{NIRdistances}).


\section{Period-Luminosity Relations\\from PHAT Photometry}\label{Results}

We construct optical PL relations using the F475W and F814W filters and NIR PL relations using the F110W and F160W filters. The measurements used in these relations are ``random phase'' magnitudes rather than time-averaged mean magnitudes. This limitation will introduce scatter into the PL relation. Although mean magnitudes lead to less scatter in the PL relations, the PHAT data alone does not have enough temporal coverage to derive mean magnitudes, unlike optimized ground-based Cepheid surveys.

\begin{figure}
\includegraphics[scale=0.45]{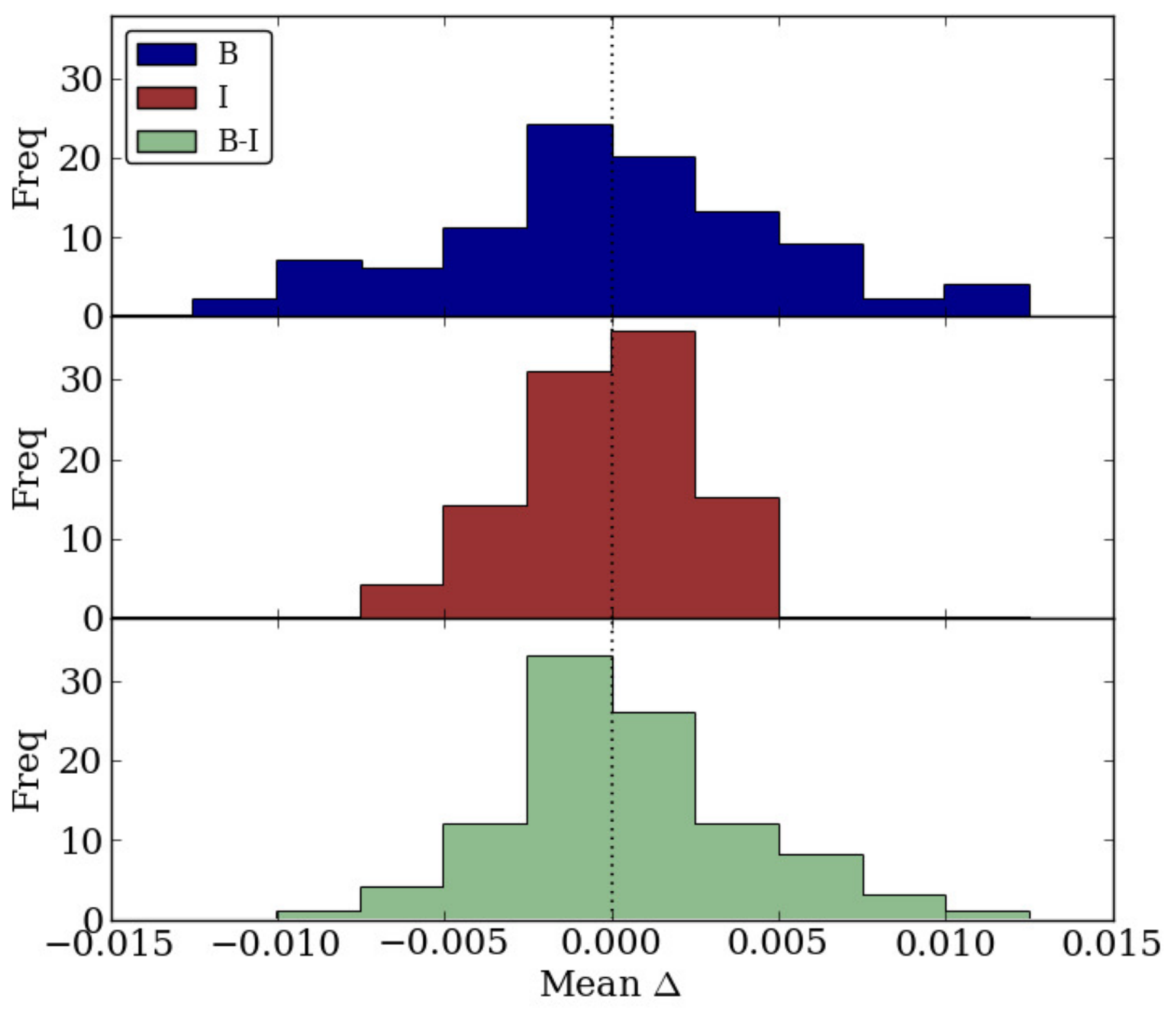}
\caption{Results of 100 Monte Carlo simulations of 165 random light curve random samples compared to the true template mean. The top panel shows the mean difference of the 165 random sample and the true magnitude for the B filter, the middle panel shows the same for the I filter, and the bottom panel shows the results   for the B$-$I color. The mean difference of the 165 random samples and the true magnitude is 0.0003 for B and -0.0001 for I; the difference for the B-I color is 0.0004. While there is scatter, there is no apparent bias in the mean.}
\label{randsamp}
\end{figure}

To determine whether we can accurately recover the mean PL relation and avoid biased distance estimates even with random phase magnitudes, we did Monte Carlo simulations of 165 light curve random samples to simulate our Cepheids using templates from Pejcha \& Kochanek (2012). We then took the mean magnitude (for both B and I) of these 165 random samples and compared it to the true mean magnitude of the template. This test was repeated 100 times and the results are shown in Figure \ref{randsamp}. The mean difference between the 165 random samples and the true magnitude is 0.0003 for B and -0.0001 for I; the difference for the B-I color is 0.0004. While there may be additional scatter around the mean due to sampling a set of Cepheids at random phase, there is no apparent bias of the mean in either magnitude or color. This suggests that random phase observations do not bias our mean distances.

\subsection{Optical Period-Luminosity Relation}\label{opticalPL}

We construct an optical Period-Luminosity relation using the random phase magnitudes from PHAT for the Cepheids in Table \ref{Cepheidlist}. We determine a Wesenheit index for the F475W and F814W filters using extinction parameters from the appendix of Schlafly \& Finkbeiner (2011) with an R$_{V}=$3.1 extinction law. The Wesenheit index takes both the brightness and color of each Cepheid into account; this index is commonly used in variable studies to correct for star-to-star differences in extinction and typically reduces the overall scatter in the P-L relation.

\begin{equation} \label{eqn hst wes}
W_{F475W,F814W}=F814W-0.879(F475W-F814W)
\end{equation}

\begin{figure*}
\includegraphics[scale=0.45]{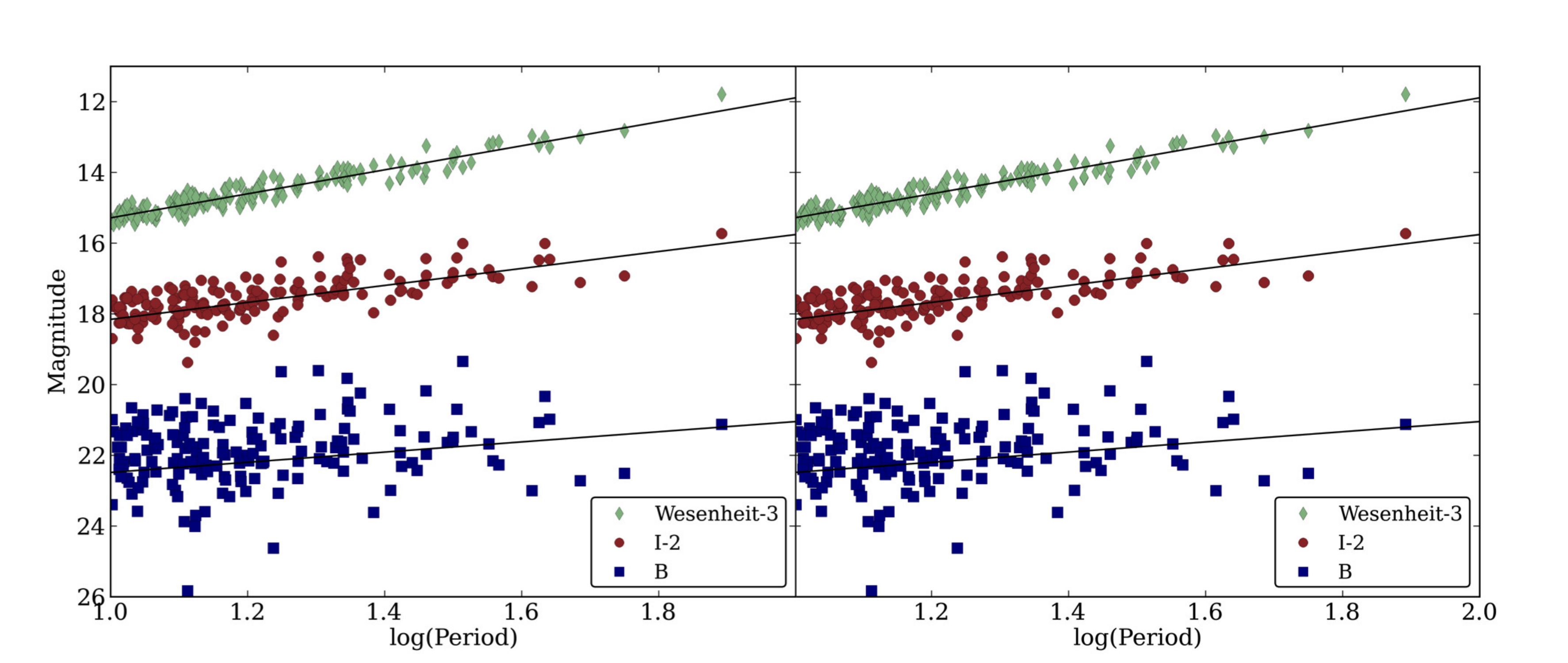}
\caption{Left panel: optical Period Luminosity relations for 163 of the 175 Cepheids with transformed  B and I magnitudes and W$_{BI}$. The I magnitudes are shifted brighter by 2 magnitudes and the Wesenheit magnitudes are shifted brighter by 3 magnitudes to make the three relations clearly visible. Right panel: optical Period Luminosity relations for 163 of the 175 Cepheids with magnitudes in F475W and F814W and W$_{F475W, F814W}$. The F814W magnitudes are shifted brighter by 2 magnitudes and the Wesenheit magnitudes are shifted brighter by 4 magnitudes to make the three relations visible. In both panels, weighted least-squares linear fits to each relation after 2.5-$\sigma$ cuts are plotted on top of the data. Uncertainties determined by artificial star tests are included, but often too small to see.}
\label{OPLplot}
\end{figure*}

We use iterative 2.5-$\sigma$ clipping with respect to the Wesenheit(F475W, F814W) relation for the 175 Cepheids with F475W and F814W magnitudes, leaving the final sample with 163 Cepheids, rejecting 7\%. We employ an iteratively re-weighted least squares, allowing the slope to float during the fitting and clipping process as opposed to clipping with respect to a fixed slope. As we sigma-clip with respect to the Wesenheit relation rather than the individual bands, we do not see a bias flattening the slope in the higher-dispersion relations. The PL relations are seen in the right panel of Figure \ref{OPLplot}. Table \ref{OPLtable} gives the slopes and intercepts of the error-weighted linear regression relations for $F475W$, $F814W$, and the Wesenheit magnitudes (W$_{F475W, F814W}$, as defined in Equation \ref{eqn hst wes}). Most notably, the dispersion in the Wesenheit(F475W, F814W) relation is only $\sim$0.17 mag, which is the lowest published dispersion for an optical PL relation in M31. Moreover, this dispersion is comparable to that found in the NIR bands (see Section \ref{IRPL}).

However, due to the lack of published PL relations in HST native filters, distance determinations are difficult. Thus, we convert the F475W and F814W magnitudes to B and I magnitudes (respectively) in order to find distances from the optical PL relations. We use the relations from Table 2 of Saha et al. (2011) to determine B and I magnitudes. We then calculate the Period-Luminosity relationship using the Wesenheit magnitude system. We adopt

\begin{equation} \label{eqn vis wes}
W_{BI}=B-1.866(B-I)
\end{equation}

\noindent from Fouqu\'e et al. (2007) with an R$_{V}=$3.1 reddening law.

Note that the color multiplier in the Wesenheit calculation is a function of the ratio of extinction in the I and B filters. However, our B magnitudes are magnitudes from the F475W passband converted to B, using semi-empirical relations that are functions of stellar temperature and typically very low extinction. To measure the impact of varying extinction on the accuracy of the Wesenheit relation using B and I magnitudes transformed from F475W and F814W magnitudes, we used Girardi et al. (2008) isochrones, based on synthetic spectra, to compare the Wesenheit magnitudes at different extinctions (A$_{V}$=0 and A$_{V}$=1) with R$_V$=3.1.

To begin, we identified points in solar-metallicity isochrones (Girardi et al. 2000, Marigo et al. 2008) falling in the instability strip in F475W and F814W magnitudes. The F475W and F814W magnitudes were converted to B and I magnitudes using the Saha et al. (2011) relations. We used the B and I magnitudes to compute Wesenheit magnitudes in the same way as our observed Cepheid magnitudes for A$_{V}$=0 and A$_{V}$=1. This process was repeated for Wesenheit magnitudes computed directly from B and I magnitudes from the same isochrone points inside the instability strip. We found that our Wesenheit BI magnitudes are not extinction-independent, but instead have a dispersion of 0.06 magnitudes between A$_{V}$=0 and A$_{V}$=1. In contrast, the same experiment done with synthetic B and I magnitudes shows a dispersion of 0.01 magnitudes. We fold an error of 0.06 into our Wesenheit magnitude uncertainty to reflect the error associated with making the filter transformation.

For the 175 Cepheids with visual magnitudes, we use a 2.5-$\sigma$ clipping with respect to the Wesenheit relation, leaving the final sample with 163 Cepheids seen in Figure \ref{OPLplot}, rejecting 7\%. As with the native filters, we allow the slope to float while fitting the PL relations and do not observe a flattening of the slope in the noisier relations. Table \ref{OPLtable} gives the slopes and intercepts of the error-weighted linear regression relations for $B$, $I$, and the Wesenheit magnitudes (W$_{BI}$, as defined in Equation \ref{eqn vis wes}). The dispersion in the Wesenheit BI relation is only 0.19 mag, which is comparable to the dispersion in the $W_{F475W, F814W}$ relation and also similar to the dispersion in the NIR filters (see Section \ref{IRPL}).

Our dispersion is less than other visual PLW (Wesenheit) relations from recent studies (e.g.: $>$0.34 mag from the DIRECT survey, 0.33 mag from Kodric et al. 2013), thereby reducing the uncertainty in the eventual distance determination. Additionally, while the sample size is smaller than many ground-based surveys, it is the largest sample size of Cepheids observed in M31 with HST in the visual bands. The high quality of HST photometry significantly reduces the dispersion seen in the PL and PLW relations, attributed primarily to the reduction of blending and crowding with HST observations. Together, these two improvements lead to gains in constraining the uncertainty in the distance modulus to M31.

We find a slope of --3.33$\pm$0.09 for the W$_{BI}$ relation, which is shallower than the slope found by Fouqu\'e et al. (2007) with Galactic Cepheids (--3.600$\pm$0.079) and their slope with LMC Cepheids (--3.454$\pm$0.011). However,  Fouqu\'e et al. (2007) includes a period range from $\log_{10}$(P) $\approx$ 0.6 to 1.8 in determining a relation, which may contribute to some discrepancy, as including shorter period Cepheids tends to steepen the PL slope.

\begin{table*}
\centering
    \caption{Optical Period-Luminosity Relations}
    \begin{tabular}{@{}cccccc@{}}
    \hline
 \textbf{Magnitude} &  \textbf{Slope} &  \textbf{Intercept (P=10d)} & \textbf{Dispersion to Fit} \\  
                               &   (mag/ Log P) & (mag)  & (mag) \\
\hline
B    &    -1.44$\pm$0.55       &	22.48$\pm$0.14	&    0.922 \\
I    &    -2.40$\pm$0.24        &	20.15$\pm$0.06	&    0.409 \\
Wesenheit (BI)   &    -3.39$\pm$0.09     & 18.29$\pm$0.02	&    0.190 \\ \hline
F475W                      &    -1.62$\pm$0.5    &    23.72$\pm$0.13    &    0.833 \\
F814W                      &    -2.41$\pm$0.24    &    22.56$\pm$0.06    &    0.403 \\
Wesenheit (F475W,F814W)    &    -3.27$\pm$0.08    &    21.83$\pm$0.02    &    0.171 \\
        \hline
    \end{tabular}
   \label{OPLtable}
\end{table*}

\subsection{Infrared Period-Luminosity Relation}\label{IRPL}

\begin{figure}
\includegraphics[scale=0.45]{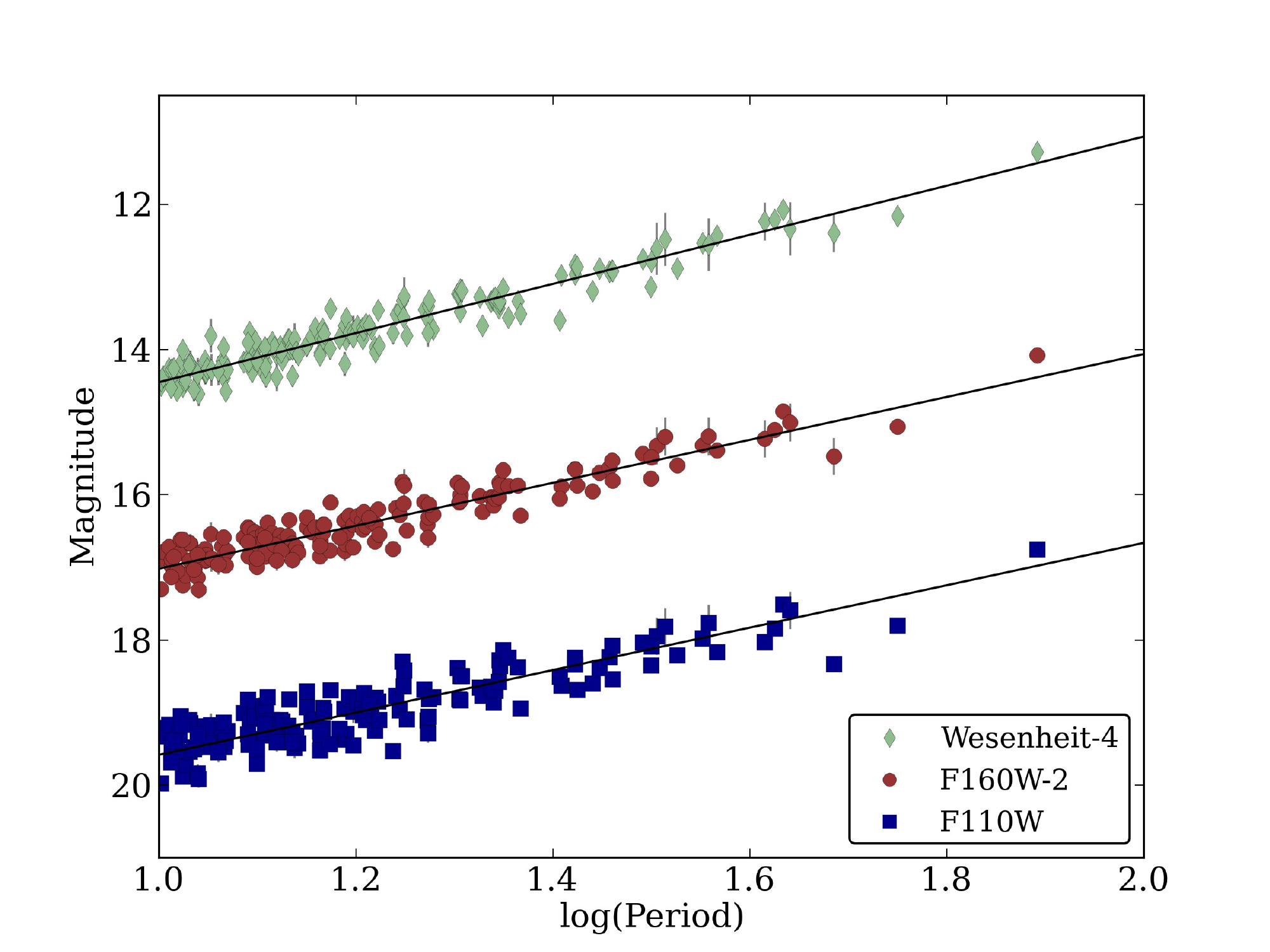}
\caption{NIR Period-Luminosity relations for 160 of the 175 Cepheids with magnitudes in F110W and F160W and W$_{IR}$. The F160W magnitudes are shifted brighter by 2 magnitudes and the Wesenheit magnitudes are shifted brighter by 4 magnitudes to make the three relations clearly visible. Weighted least-squares linear fits to each relation after 2.5-$\sigma$ cuts are plotted on top of the data. Uncertainties determined by artificial star tests are included, though for some points they are too small to see clearly.}
\label{IRPLplot}
\end{figure}

In Figure \ref{IRPLplot}, we show the Period-Luminosity relation from the random phase magnitudes for the Cepheids in F110W and F160W WFC3 filters, and in Wesenheit magnitudes, as calculated as in Riess et al. (2012) with an R$_{V}=$3.1 reddening law:

\begin{equation} \label{Wes_IR}
W_{IR}= F160W - 1.54(F110W - F160W)
\end{equation}
\noindent

As with the optical bands, we fit linear, error-weighted relations to the data. The resulting slopes, intercepts, and dispersions are given in Table \ref{IRPLtable}. Out of the 175 Cepheids with IR magnitudes, 160 are left after 2.5-$\sigma$ clipping with respect to the Wesenheit relation (allowing the slope to float, producing a 9\% rejection). Although we present the NIR Period-Luminosity relations in Table \ref{IRPLtable}, we do not directly use these fits to determine a distance in Section \ref{NIRdistances}. As expected, the individual NIR bands give a much smaller scatter in the Period-Luminosity relation, due to lower extinction and smaller overall amplitudes of the Cepheid light curves. However, although the amplitudes in the NIR are generally less than those in the visual, they are not completely negligible (Monson \& Pierce 2011, Testa et al. 2007, Persson \& Madore 2004, Welch et al. 1984).

The dispersion in the Wesenheit relation is similar to previous studies and comparable to that of the visual Wesenheit dispersion. Our results show slightly higher dispersion in F110W (0.246 mag) than found by Riess et al. (2012, 0.20 mag), but very similar to that of Kodric et al. (2015, 0.243). It is possible the discrepancy in the dispersions between our study and Riess et al. (2012) is related to the differences in photometry. Similar dispersions in the F160W PL are found in all three studies: 0.187 mag in this paper; 0.17 mag in Riess et al. 2012; 0.178 in Kodric et al. (2015). We find a dispersion of 0.173 mag in the PLW relation, between 0.22 from Riess et al. (2012) and 0.147 Kodric et al. (2015), although we note that Kodric et al. use a different definition of the Wesenheit magnitude. Save for F110W, our dispersions are also consistent with the expected random phase dispersions as noted by Riess et al. (2012) from the study of Persson et al. (2004) of LMC Cepheids.

Generally, the IR bands provide an opportunity to improve upon the dispersion of the PL relation typically seen in the visual bands. However, we do not necessarily see a significant difference between the dispersions of PLW relations in the visual and in the IR, although the individual bands show immense improvement compared to the visual as the wavelength gets longer, as expected. By increasing the sample size of long-period Cepheids in the IR bands over Riess et al. (2012, 68 Cepheids) and Kodric et al. (2015, 110 Cepheids), we have further reduced the statistical uncertainty in the distance measurement.

\begin{table*}
\centering
    \caption{Infrared Period-Luminosity Relations}
    \begin{tabular}{@{}cccccc@{}}
    \hline
 \textbf{Magnitude} &  \textbf{Slope}   &\textbf{Intercept (P=10d)}  & \textbf{Dispersion to Fit} \\  
                                &   (mag/ Log P) & (mag)  & (mag) \\
\hline
F110W    &    -2.92$\pm$0.11        &	19.58$\pm$0.03	&    0.246 \\
F160W    &    -2.96$\pm$0.09       &	 19.02$\pm$0.02   &    0.187 \\
Wesenheit    &    -3.38$\pm$0.08         &	 18.45$\pm$0.02  &    0.173 \\
        \hline
    \end{tabular}
   \label{IRPLtable}
\end{table*}

Kodric:
F110W: --2.497$\pm$0.209
F160W: --2.779$\pm$0.171

Comparing the linear fits of the F160W filter from Table \ref{IRPL} to Kodric et al. (2015), we find a similar slope (--2.96$\pm$0.09 from this paper and --2.779$\pm$0.171) and intercept (19.02$\pm$0.02 from this paper and 18.960$\pm$0.028 from Kodric et al. 2015). The F110W slope from Kodric et al. (2015) of --2.497$\pm$0.209 is significantly flatter than the slope we find of --2.92$\pm$0.11. The intercepts are reasonably similar within 1.5-$\sigma$ (19.58$\pm$0.03 from this paper and 19.476$\pm$0.037 from Kodric et al. 2015).

We find that our IR Wesenheit slope of --3.38$\pm$0.08 is consistent with Riess et al. (2012), who found a slope of $-3.43\pm$0.17 and with Persson et al. (2004) for the LMC, who found a slope of --3.38$\pm$0.09. The period range of Persson et al. (2004) (extending to $\log_{10}$(P)$\approx$2) is comparable to that of Riess et al. (2012) and our own complete sample in the NIR (extending to $\log_{10}$(P)$\approx$1.9, see Figures \ref{periods} and \ref{IRPLplot}). Our slope is steeper than that of Kodric et al. (2015) at --3.172$\pm$0.117; however, they calculate the Wesenheit magnitude slightly differently so the comparison is incomplete.


\section{Ground vs. HST PL Relations}\label{DIRECTPL}

We follow a similar procedure as Section \ref{opticalPL} to analyze the subset of DIRECT Cepheids in the PHAT photometric catalog. From this, a comparison of ground-based and HST photometry can be made. We use the PHAT photometry and published DIRECT periods and magnitudes to analyze the differences between space-based random phase magnitudes and ground-based mean magnitudes. For each of the 80 Cepheids in Table \ref{DIRECTCephs}, we proceed as follows. First, we note that, because the DIRECT survey is mainly in the $V$ and $I$ (only about half the Cepheids have $B$ magnitudes) bands, the $V$-band and $I$-band magnitudes must be inferred from the HST data to make a comparison of the PHAT survey to the DIRECT survey.

To transform F814W magnitudes to $I$ magnitudes, we use the Saha et al. (2011) relations. We determine a color transformation from the F475W and F814W filters to the $V$-band Johnson-Cousin filter through the use of Girardi isochrones at a variety of ages (1 Gyr through 10 Gyr) and extinctions (A$_{V}$ from 0 to 2 mags). The transformation is restricted to the F475W--F814W color range of our DIRECT sample (F475W--F814W from 0.9 to 3.2). The conversion we determine is:

\begin{equation}\label{Vband}\begin{split}
V- F814W\approx 0.104+0.541(F475W-F814W)\\
+0.032(F475W-F814W)^2
\end{split}\end{equation}
\noindent

\noindent with an dispersion of 0.03 mags in the transformation. The PHAT transformed $V$ and $I$ magnitudes are used to construct PL relations and compared to the PL relations using the published $V$ and $I$ magnitudes from the DIRECT survey.

We calculate Wesenheit magnitudes for each Cepheid via the following relation from Fouqu\'e et al. (2007), where W$_{V}$ is the Wesenheit magnitude, $V$ is as in Equation \ref{Vband}, and $I$ is transformed via Saha et al. (2011) from F814W:

\begin{equation} \label{Wes_optical}
W_{VI} = V - 2.55(V - I)
\end{equation}
\noindent

\begin{figure*}
\includegraphics[scale=0.45]{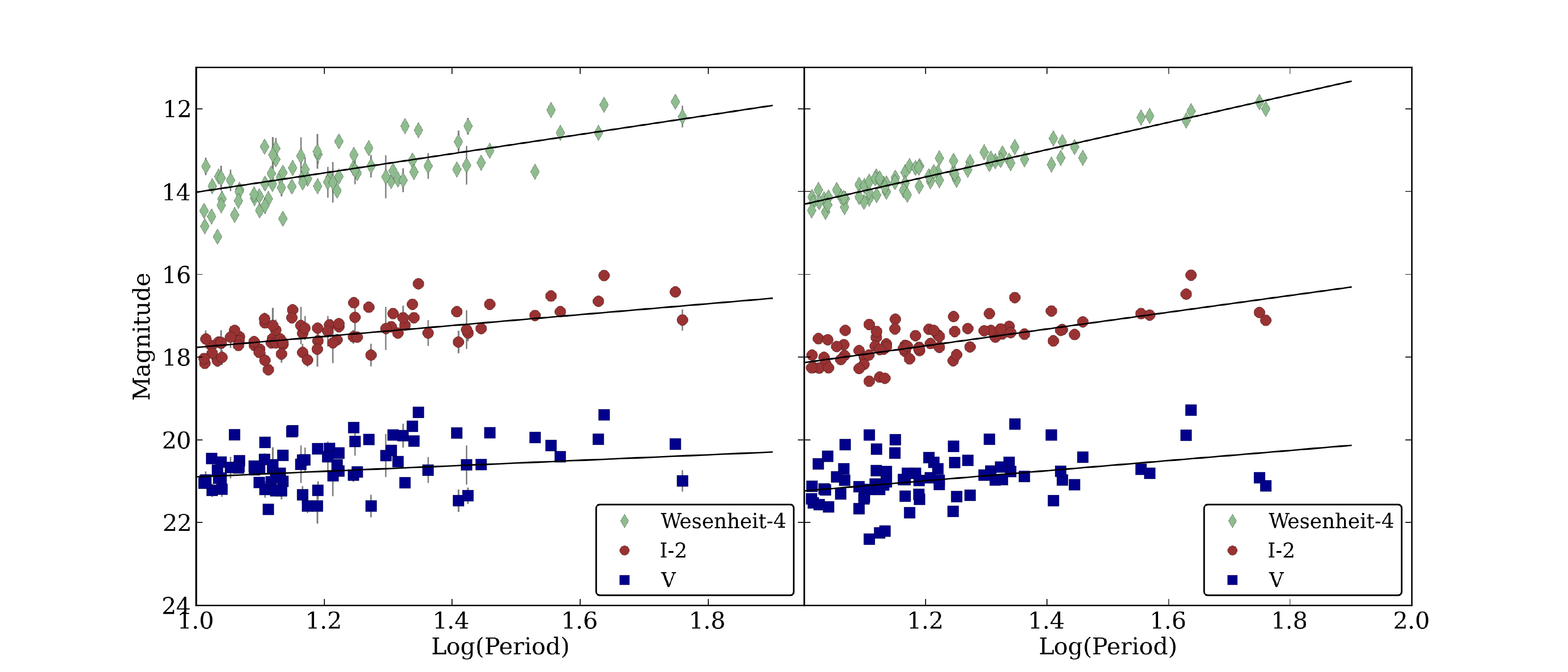}
\caption{Optical PL relations for ground-based DIRECT (left), for random phase HST (right) DIRECT Cepheids with V, I (transformed from F475W and F814W), and W$_{VI}$ magnitudes. The I magnitudes are shifted brighter by 2 magnitudes and the Wesenheit magnitudes are shifted brighter by 4 magnitudes to make the three relations clearly visible in each panel. The left panel includes the published DIRECT ground-based mean magnitudes for each Cepheid, with 77 Cepheids and error bars as given in the published DIRECT data. The right panel shows the same Cepheids' random phase magnitudes determined from the PHAT photometry, with 75 Cepheids and error bars as determined from artificial star tests. Weighted least-squares linear fits to each relation after 2.5-$\sigma$ cuts are plotted on top of the data in both panels.}
\label{direct}
\end{figure*}

The PHAT PL relations for $V$, $I$, and Wesenheit magnitudes are shown in the right panel of Figure \ref{direct}. For comparison, PL relations from the DIRECT ground-based mean magnitudes are shown in the left panel of Figure \ref{direct}. We use 2.5 $\sigma$-clipping with respect to the Wesenheit magnitudes and allow the slope to float. Out of the 80 Cepheids in the sample, this leaves 77 Cepheids in the DIRECT photometry (a 4\% rejection rate) and 75 Cepheids in the PHAT photometry (a 6\% rejection rate). Along with their dispersions, the linear, error-weighted fits to the relations in $V$, $I$, and Wesenheit magnitudes from Figure \ref{direct} are given in Table \ref{directtable}. 

\begin{table*}
\centering
    \caption{DIRECT Ground-Based Period-Luminosity Relations}
    \begin{tabular}{@{}cccccc@{}}
    \hline
  & \textbf{Magnitude} &  \textbf{Slope} & \textbf{Intercept (P=10d)} & \textbf{Dispersion to Fit} \\  
                                 &                &   (mag/ Log P) & (mag)  & (mag) \\
\hline
DIRECT Ground-Based   & V         & -0.67$\pm$0.33  & 20.89$\pm$0.09 & 0.508 \\
                  & I           & -1.32$\pm$0.21   & 19.76$\pm$0.06	& 0.339 \\
                  & Wesenheit & -2.33$\pm$0.30 & 18.01$\pm$0.08 & 0.440 \\
PHAT Random Phase & V         & -1.22$\pm$0.44  &	21.23$\pm$0.11 & 0.541 \\
                  & I         & -2.03$\pm$0.24  &  20.13$\pm$0.06 & 0.327 \\
                  & Wesenheit & -3.30$\pm$0.12 &  18.31$\pm$0.03 & 0.187 \\
                  \hline
    \end{tabular}
   \label{directtable}
\end{table*}

\begin{figure*}
\includegraphics[scale=0.45]{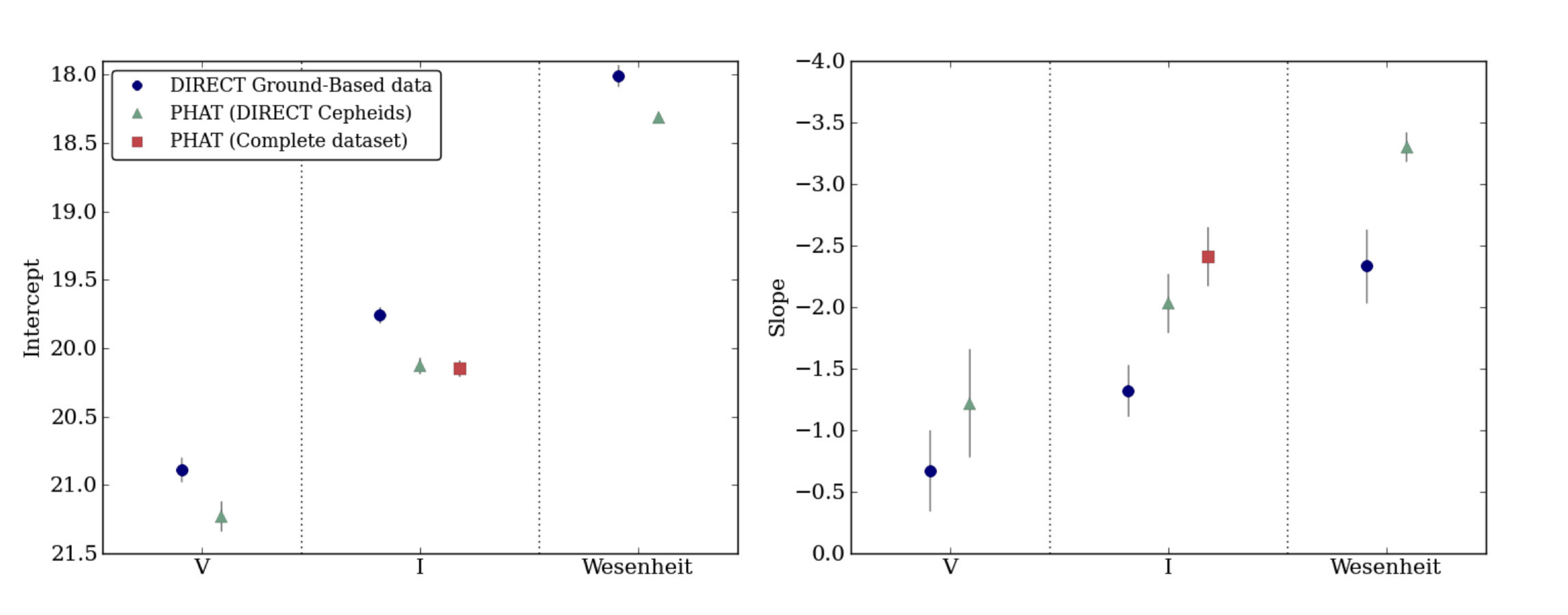}
\caption{Left panel: a comparison of intercept values across several filters; circles denote intercepts from the DIRECT ground-based PL relation, triangles indicate the intercepts from the DIRECT sample as observed by PHAT, and squares indicate the intercepts from the complete PHAT sample ($I$ band only). The left section shows intercepts from the $V$-band PL relation, the middle from the $I$-band, and the right section from the W$_{VI}$ PL relation. Error bars are also plot for each intercept. Right panel: a comparison of slope values across several filters; circles denote slopes from the DIRECT ground-based PL relation, triangles indicate the slopes from the DIRECT sample as observed by PHAT, and squares indicate the slopes from the complete PHAT sample ($I$ band only). The left section shows slopes from the $V$-band PL relation, the middle from the $I$-band, and the right section from the W$_{VI}$ PL relation. Error bars are also plot for each slope.}
\label{slpint}
\end{figure*}

As illustrated in Figure \ref{slpint}, the ground-based DIRECT Period-Luminosity relations have much shallower slopes and brighter intercepts compared to the HST PL relations. A shallower slope leads to a longer distance modulus; however, a brighter intercept leads to a shorter distance modulus. These two effects may balance out in distance determinations, but it is evident that ground-based observations are susceptible to heavily biased slopes and intercepts and caution should be taken when drawing conclusions from ground-based PL relations. This behavior suggests that blending and crowding may be the most significant limitation for ground-based Cepheid studies; these issues are improved by the spatial resolution of ACS/WFC. We note, however, that the level of crowding in ground-based surveys of M31 is comparable to that expected for HST surveys of more distance galaxies. Therefore, we expect many HST Cepheid studies in more distant galaxies may be affected by the same biases affecting the DIRECT observations of M31 (Stanek \& Udalski 1999, Mochejska et al. 2000, Chavez et al. 2012).

As Figure \ref{slpint} shows, the difference between the slopes and the intercepts between the ground-based PL relations and the PHAT PL relations grows as we move from the V band to the I band. The difference is most stark when comparing the Wesenheit slopes and intercepts. This implies that the ground-based bias significantly affects both the slope and the intercept of the PL and PLW relations, thereby jeopardizing accurate distance estimates. The existence of this bias stresses the need for more HST observations of Cepheids, even if they are only at random phases.

The slope of the visual Wesenheit VI relation from the PHAT data is in agreement with previous studies. We find a slope of --3.30$\pm$0.12, comparable to that found by Benedict et al. (2007) and Fouqu\'e et al. (2007) for Galactic Cepheids (--3.31$\pm$0.17 and --3.377$\pm$0.023) and that of Udalski et al. (1999) for the LMC (--3.28$\pm$0.14), although the period ranges differ. The periods in Benedict et al. (2007) range from very low period ($\log_{10}$(P)=0.4) to longer periods ($\log_{10}$(P)=1.6), but the data points are extremely sparse between $\log_{10}$(P)=1 and their longest period Cepheid at $\log_{10}$(P)=1.6. The sample in Fouqu\'e et al. (2007) is better populated in period, with a range from approximately $\log_{10}$(P)=0.6 to $\log_{10}$(P)=1.7. For the Cepheids in the complete dataset, the period coverage is out to $\log_{10}$(P)=1.9, a bit farther than Fouqu\'e et al. (2007). However, we do not include the short period ($<$10 days) Cepheids in our determination of the PL relations.


\section{Distance moduli}\label{dists}

\subsection{Distances: Visual}

To determine a distance modulus for the complete dataset in the visual bands, we use the Wesenheit BI relation as defined by Fouqu\'e et al. (2007), calibrated to Galactic Cepheids. The distance modulus in these filters is defined by:

\begin{equation}\label{FBI}
M_{W(BI)}=-3.600(\pm0.079)\log_{10}(P)-2.401(\pm0.023)
\end{equation}
\noindent

\noindent and

\begin{equation} \label{foqueBI}
\mu_{M31_{BI}}=m_{W(BI)}-M_{W(BI)}
\end{equation}
\noindent

\noindent where M$_{W(BI)}$ is the magnitude of the Wesenheit BI relation from Fouqu\'e et al. (2007) evaluated at $\log_{10}$(P) of 1.26, the mean period of our sample, and m$_{W(BI)}$ is the magnitude of the Wesenheit BI from our observations, also evaluated at $\log_{10}$(P)=1.26. The result is a distance modulus of 24.34$\pm$0.10.

We also determine a distance modulus using the LMC-based relation from Fouqu\'e et al. (2007), using the same formalism as Equations \ref{FBI} and \ref{foqueBI}, except with a slope of -3.454$\pm$0.011 and an intercept of 15.928$\pm$0.003 (adopting a distance to the LMC of 18.486$\pm$0.065 as in Riess et al. 2011). The result is a distance modulus of 24.31$\pm$0.07. We average our two visual band distance determinations to obtain a modulus of \VISdist{}.

\subsection{Distances: IR}\label{NIRdistances}

To obtain a distance modulus to M31 for the NIR Cepheids, we tie the distance modulus of M31 to that of NGC 4258, as defined in Riess et al. (2009b) and Riess et al. (2011) (revisited by Riess et al. 2012 and Fiorentino et al. 2013). From Equation 7 of Riess et al. (2009b), we solve for the distance modulus to M31 as 

\begin{equation}\label{distmodeq}\begin{split}
\mu_{M31} = \mu_{4258}-zp_{4258}+m_{W}-b_{W}\log_{10}(P-1)\\
+Z_{W}\Delta \log_{10}[O/H]
\end{split}\end{equation}
\noindent

\noindent where $\mu_{4258}$ is the distance to the water maser galaxy NGC 4258 (7.6 $\pm$ 0.17 $\pm$ 0.15 Mpc from Humphreys et al. 2013), $zp_{4258}$ is the intercept of the PL relation for Cepheids in NGC 4258, $m_{W}$ are the observed NIR Wesenheit magnitudes of the Cepheids, $b_{W}$ is the slope of the global Wesenheit NIR PL relation, and $P$ are the periods (\textgreater 10 days). The term $Z_{W}\Delta \log_{10}[O/H]$ is the metallicity dependence term; $\Delta \log_{10}[O/H]$ are the metallicities of the Cepheids relative to the LMC and $Z_{W}$ is the global metallicity slope. To obtain the values of $zp_{4258}$, $b_{W}$, and $Z_{W}$ we use a simultaneous linear fit to the Cepheids in both M31 and NGC 4258. The simultaneous linear fit allows a reduction in the final distance uncertainty compared to using the NIR PL relations alone.

First, we should note that same filters were not used to observe Cepheid colors in NGC 4258 (V--I) as the bands observed in the PHAT survey (F160W--F110W). The Wesenheit magnitudes in Riess et al. (2009 and 2011) were calculated using W$_{IR}$=F160W--1.54*(V--I). To account for the difference in defining the Wesenheit magnitude, a correction has been applied to the (F110W--F160W) based extinction correction to give them the same mean color in (V--I) as M31 Cepheids. As in Riess et al. (2012), we use the following to determine a correction (X) to the Wesenheit magnitude calculation of NGC 4258:

\begin{equation}\label{correction}
0.504(V - I) = 1.54(F110W - F160W - X)
\end{equation}
\noindent

With a mean (V--I) color of 1.23 (from the DIRECT Cepheids) and a mean (F110W--F160W) color of 0.56 from the PHAT photometry, we obtain a value of X=0.156. Because of the difference in F110W photometry between PHAT and Riess et al. (2012), we use a different X than Riess et al. (2012) to place our Wesenheit magnitudes on the same scale as that of the NGC 4258 photometry. Errors in their NGC 4258 photometry are incorporated into the analysis and error budget.

To solve for the distance, we also must determine the values of $\Delta \log_{10}[O/H]$ for the Cepheids to include the possibility that differing metallicities of Cepheids could introduce scatter into the P-L relation. To examine possible metallicity effects, we use the relation from Zaritsky et al. (1994) for M31 to determine $\log_{10}$[O/H] for each Cepheid from its radial location in the disk and compare it to the solar value of 8.9 (as in Riess et al. 2012, Riess et al. 2011, and Riess et al. 2009). 

We adopt a center for M31 of RA = 0$^{h}$ 42$^{m}$ 44.31$^{s}$ and Dec = 41$^{\circ}$ 16' 9.4" (Cotton et al. 1999), a disk scale length of $\rho_{o}$=77.44 (Zaritsky et al. 1994), and position angle $\phi$=37.715 (Haud et al. 1981, Baade \& Arp 1964). These values are used to calculate the 12+$\log_{10}$[O/H] metallicity term from the relation determined by Zaritsky et al. (1994), as seen in Equation \ref{zaritsky}.

\begin{equation} \label{zaritsky}\begin{split}
12+\log_{10}([O/H])=\\
9.03(\pm0.09)-0.28(\pm0.10)(\rho/\rho_{0}-0.4)
\end{split}\end{equation}
\noindent

The deprojected radius, $\rho$, is calculated using Equations \ref{xeq} through \ref{rhoeq}:

\begin{equation} \label{xeq}
x=(\delta-\delta_{0})cos(\phi)+(\alpha-\alpha_{0})sin(\phi)cos(\delta_{0})
\end{equation}
\noindent

\begin{equation} \label{yeq}
y=\frac{(\delta-\delta_{0})sin(\phi)-(\alpha-\alpha_{0})cos(\phi)cos(\delta_{0})}{cos(i)}
\end{equation}
\noindent

\begin{equation} \label{rhoeq}
\rho=\frac{(x^{2}+y^{2})^{^{\frac{1}{2}}}}{\rho_{0}}
\end{equation}
\noindent

\begin{figure}
\includegraphics[scale=0.45]{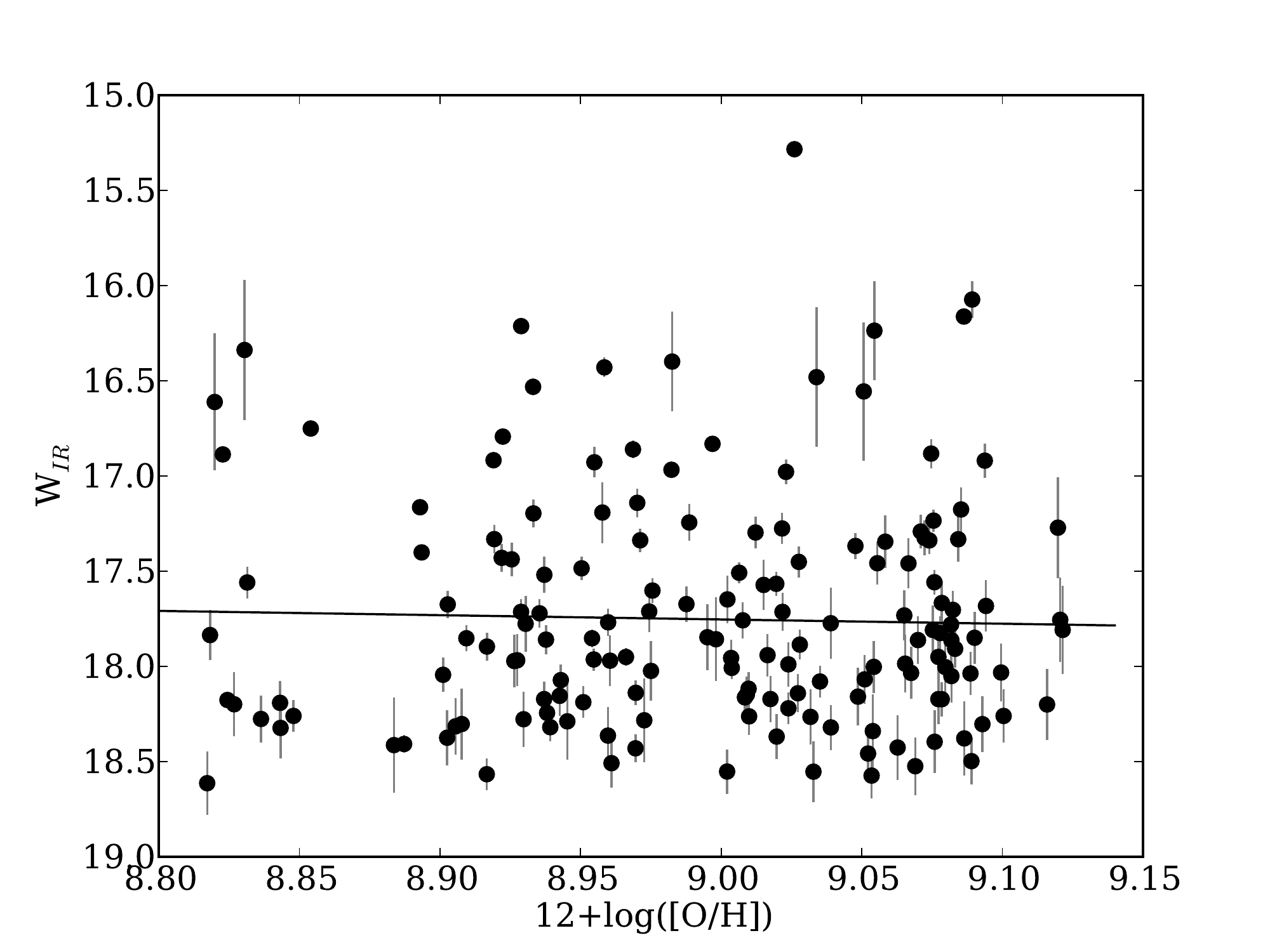}
\caption{The 12+$\log_{10}$[O/H] vs $W_{IR}$ magnitude relation for 160 Cepheids in M31. The black points indicate the individual Cepheids' metallicities determined by the relation from Zaritsky et al. (1994), Equation \ref{zaritsky}, with propagated errors. The solid line shows the linear, error-weighted fit to the Cepheids, with a slope of 0.22$\pm$0.60 mag/dex, which is not statistically significant at the 0.05 level with a two-sided student t-test value of 0.37 and 159 degrees of freedom. We see a range of $\sim$0.3 dex in 12+$\log_{10}$[O/H] from approximately 8.82 to 9.12.}
\label{logOH}
\end{figure}

The resulting 12+$\log_{10}$[O/H] values are compared to the IR Wesenheit magnitude in Figure \ref{logOH}. The slope of a linear fit relating the metallicity of the Cepheids to their magnitudes, is 0.22$\pm$0.60 mag dex$^{-1}$. A two-sided student t-test with a value of t=0.37 and 159 degrees of freedom informs that this is not a statistically significant relation at the 0.05 significance level. Our slope of 0.22$\pm$0.60 mag dex$^{-1}$ is different than Riess et al. (2012), who obtained a slope of --0.65$\pm$0.73, but with large uncertainty.

We see a range of $\sim$0.3 dex in 12+$\log_{10}$[O/H] from approximately 8.82 to 9.12, similar to Riess et al. (2012), who find a range of 8.87 to 9.05. This range is similar to the $\sim$0.4 dex that we expect from the --0.018 dex/kpc gradient of Zaritsky et al. (1994) over the range of the disk. The Cepheids in M31 do not have a clear magnitude-metallicity dependence, as has been found in previous studies (Riess et al. 2012, Freedman \& Madore 1990). However, the values of 12+$\log_{10}$[O/H] for M31 Cepheids will be used in conjunction with those in NGC 4258 to find a global fit.

Using multiple error-weighted linear regression, we simultaneously fit a period-luminosity relation to the Cepheids in M31 and those in NGC 4258 to determine values for $zp_{4258}$, $b_{W}$, and $Z_{W}$. These values are then used to determine the distance modulus to M31 as in equation \ref{distmodeq}. We determine a $b_{W}$ slope of --3.24$\pm$0.10, $zp_{4258}$ of 23.26$\pm$0.06, and $Z_{W}$ of 0.16$\pm$0.26. These are comparable to the values determined by the global fit in Riess et al. (2011, see Table \ref{globalfit}). Note that while Riess et al. (2011) use all SN hosts for a global fit, we use NGC 4258 with M31 only. With the values in Table \ref{globalfit}, we obtain a distance of \NIRdist{} to M31.

\begin{figure}
\includegraphics[scale=0.45]{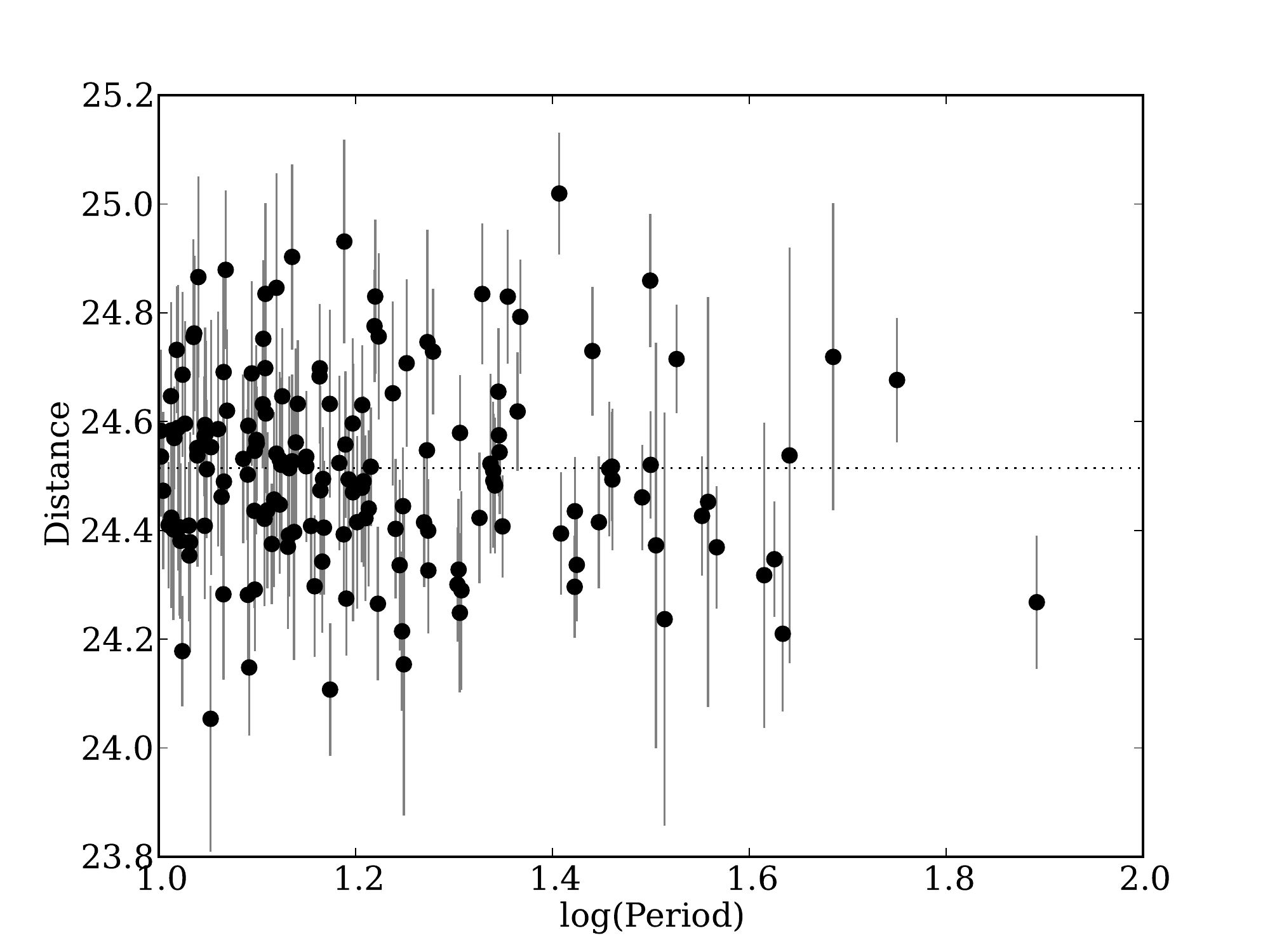}
\caption{Individual IR distances (described in Section \ref{NIRdistances}) with propagated errors are plotted against period. No correlation between the distance modulus and the period of variability is seen. The dashed line indicates the mean distance.}
\label{IRdistplot}
\end{figure}

The derived distance moduli for the complete dataset in the visual (with MWG and LMC anchors) and in the IR are given in Table \ref{distmods}. There is an offset between the visual and the IR distances. We discuss this difference in Section \ref{Discussion} below. We plot the individually determined distances derived from the IR photometry against period in Figure \ref{IRdistplot}; there does not to appear to be a trend with period. Published period uncertainties are not available for the DIRECT data and the POMME Cepheids from Riess et al. (2012), so we cannot fully incorporate period uncertainties into our analysis. However, if we adopt a reasonably conservative 0.1 day period error, the uncertainty in the NIR and visual distances is affected by less than 0.01.

\begin{table}
\centering
    \caption{Global Fit Values}
	\begin{tabular}{@{}lll@{}}
    \hline
 \textbf{Variable} (Eq.\ref{distmodeq}) &  \textbf{This Paper} & \textbf{Riess et al. 2011} \\  
\hline
$\mu_{4258}$		&	7.6$\pm$0.17$\pm$0.15	&	7.2$\pm$0.2$\pm$0.3	\\
zp$_{4258}$		&	23.26$\pm$0.06		&	26.32$\pm$0.03   (23.10 at P=10 d)	\\
b$_{W}$			&	-3.24$\pm$0.10		&	-3.21$\pm$0.03	\\
Z$_{W}$			&	0.16$\pm$0.26			&	-0.10$\pm$0.09	\\
        \hline
    \end{tabular}
   \label{globalfit}
\end{table}

\begin{table*}
\centering
    \caption{Distance Moduli}
    \begin{tabular}{@{}cccc@{}}
    \hline
\textbf{Dataset Sample} &  \textbf{Filter} & \textbf{Anchor} & \textbf{$\mu$} \\  
\hline
Complete 	& PHAT Visual (B, I) 			&	Galactic Cepheids	& 24.34$\pm$010 \\
			&				 			&	LMC				& 24.31$\pm$0.07 \\
         		& PHAT NIR (F110W, F160W)	&	NGC 4258			& \NIRdist{} \\
\hline
DIRECT	 	& Ground-based (V, I) 		&	Galactic Cepheids	& 24.21$\pm$0.11 \\
		 	& 					 		&	LMC				& 24.21$\pm$0.09 \\
       			& PHAT Visual (V, I)			&	Galactic Cepheids	& 24.24$\pm$0.10 \\
			&				 			&	LMC				& 24.24$\pm$0.08 \\
\hline
    \end{tabular}
   \label{distmods}
\end{table*}

\subsection{Ground vs. HST Distances}

For the DIRECT Cepheid dataset, we use the Wesenheit VI relation as defined by Fouqu\'e et al. (2007), calibrated to Galactic Cepheids. We obtain distance moduli for the random phase PHAT Cepheid PL Wesenheit relations and the ground-based DIRECT PL Wesenheit relations in the $V$ and $I$ bands by comparing the magnitude of our PL relation to the calibrated absolute magnitude PL relation at the mean period of our sample (P=1.26). The absolute Wesenheit magnitude is given by Fouqu\'e et al. (2007) as:

\begin{equation}\label{FVI}
M_{W(VI)}=-3.477(\pm0.074)\log_{10}(P)-2.414(\pm0.022)
\end{equation}
\noindent

The distance modulus is then defined as:

\begin{equation} \label{foqueVI}
\mu_{M31_{VI}}=m_{W(VI)}-M_{W(VI)}
\end{equation}
\noindent

\noindent where M$_{W(VI)}$ is the magnitude of the Wesenheit VI relation from Fouqu\'e evaluated at $\log_{10}$(P) of 1.26 (the mean period of our sample) and m$_{W(VI)}$ is the magnitude of the Wesenheit VI relation from our observations, also evaluated at $\log_{10}$(P)=1.26.
We also determine a distance modulus using the relation form Udalski et al. (1999), using the same formalism as Equations \ref{FVI} and \ref{foqueVI}, with a slope of --3.320$\pm$0.014 and an intercept of 15.880$\pm$0.010 (adopting an LMC distance modulus of 18.486$\pm$0.065 as in Riess et al. 2011).

Our W(VI) DIRECT sample extends to a maximum period of about $\log_{10}$(P)=1.75, comparable to that of Fouqu\'e et al. (2007). Udalski et al. (1999) have period coverage from $\log_{10}$(P)=0.4 to $\log_{10}$(P)=1.5, where the LMC Cepheids begin to saturate their detector. In both cases, the authors include the shorter period Cepheids in determining a PL relation; however, we restrict our sample to only those Cepheids with periods greater than 10 days.

Distance estimates for the DIRECT subset from both the ground-based DIRECT photometry and the PHAT photometry are included in Table \ref{distmods}. Distances are given for both a Milky Way and an LMC anchor. 

The ground-based distance moduli appear to be shorter than their HST counterparts by $\sim$0.03. While this difference is well within the error, it could also be due to less accurate photometry on the ground. Blending and crowding could plausibly cause systematically shorter distances from ground-based photometry. We discuss this issue further in section \ref{Discussion}.


\section{Metallicity and the Distance-Radius Relation}\label{Distgrad}

A slight trend of Cepheid distance modulus with radial location from the center of the galaxy has been seen in M101 and M81, among others (see Freedman \& Madore 1990, Kennicutt et al. 1998, Macri et al. 2006, McCommas et al. 2009, Gerke et al. 2011). The discrepancy in distance between Cepheids in the inner and outer regions of these galaxies is often attributed, at least in part, to metallicity effects on the Period-Luminosity Relation, although the cause is still debated (Gould 1994; Vilardell et al. 2007, Gerke et al. 2011, Majaess et al. 2011, Kudritzki et al. 2012).

\begin{figure}
\includegraphics[scale=0.45]{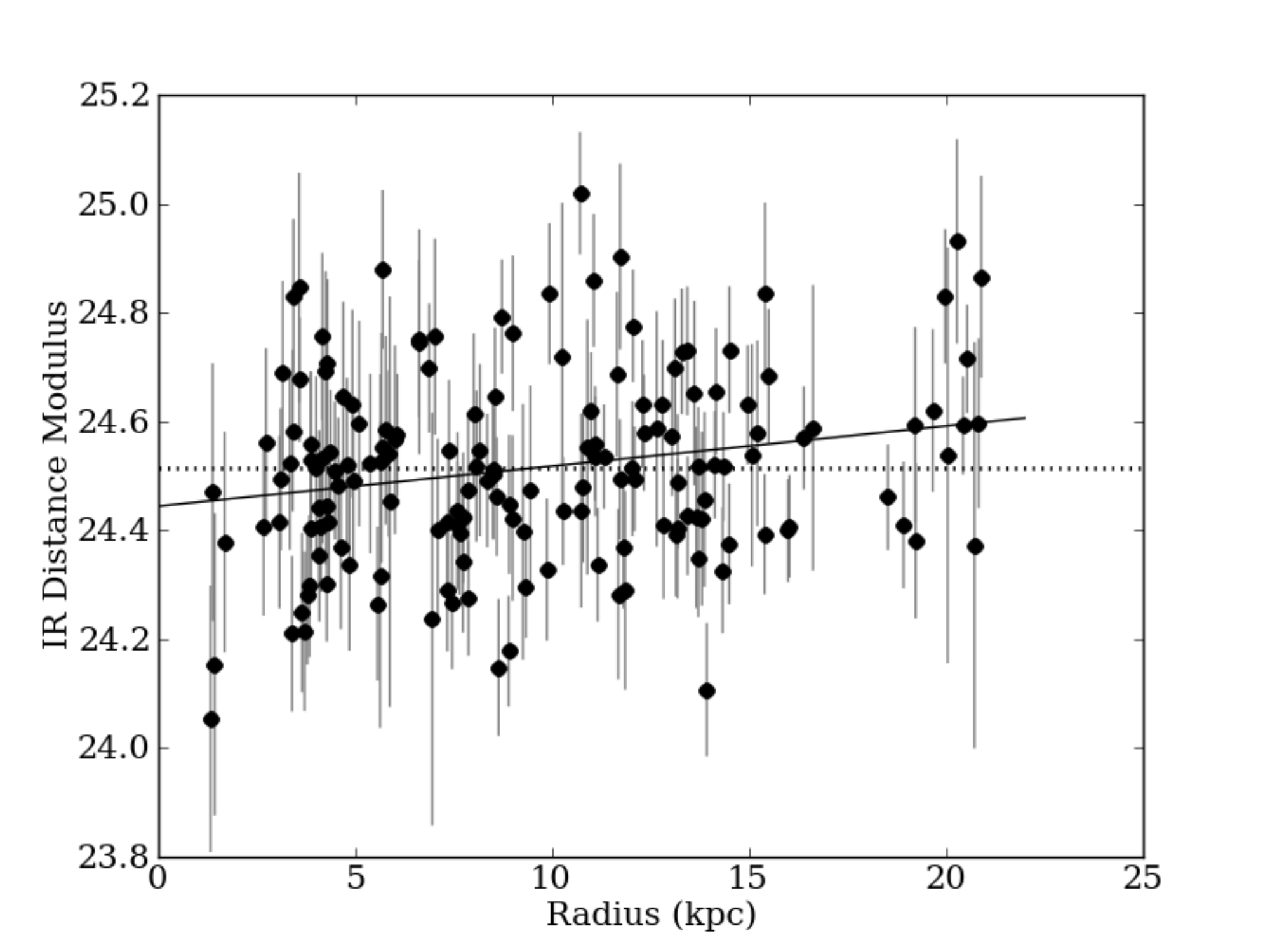}
\caption{The Distance-Radius Relation for the 160 Cepheids in M31 that were used in the final PL relations in Section \ref{IRPL} using the complete dataset. The black points indicate the individual Cepheids' IR distances with propagated errors. The horizontal dotted line shows the mean distance determination from Section \ref{dists}, and the solid line shows the linear fit to the data. The linear fit has a slope of 0.007 $\pm$ 0.003, and a two-sided student t-test value of 2.71 (with 159 degrees of freedom), making it significant at the 99.5\% level.}
\label{dist-rad}
\end{figure}

We use our complete set of Cepheids from Table \ref{Cepheidlist} to explore the relationship between distance and radius in M31, determining the radius of each Cepheid by using the right ascension and declination to determine its de-projected position in the galaxy, $\rho$, as seen in Equation \ref{rhoeq}. We convert $\rho$ to kpc using the mean IR distance modulus determined in Section \ref{dists}. The individual distances in the IR bands ($\mu_{IR}$) for each Cepheid are compared to the de-projected radii in Figure \ref{dist-rad}. A linear fit yields a gradient of 0.007$\pm$0.003. The two-sided student t-test value for this relation is 2.71, making it significant beyond the 99.5\% level with 159 degrees of freedom. We apply Chauvenet's criterion to the data to determine the sensitivity of individual points on the relation. We find 6 outlying points with a significance level of 0.05. While this shallows the slope slightly, the slope remains statistically significant. This suggests that there is a statistical difference in the distance moduli of individual Cepheids in the inner and outer portions of M31.

\begin{figure}
\includegraphics[scale=0.45]{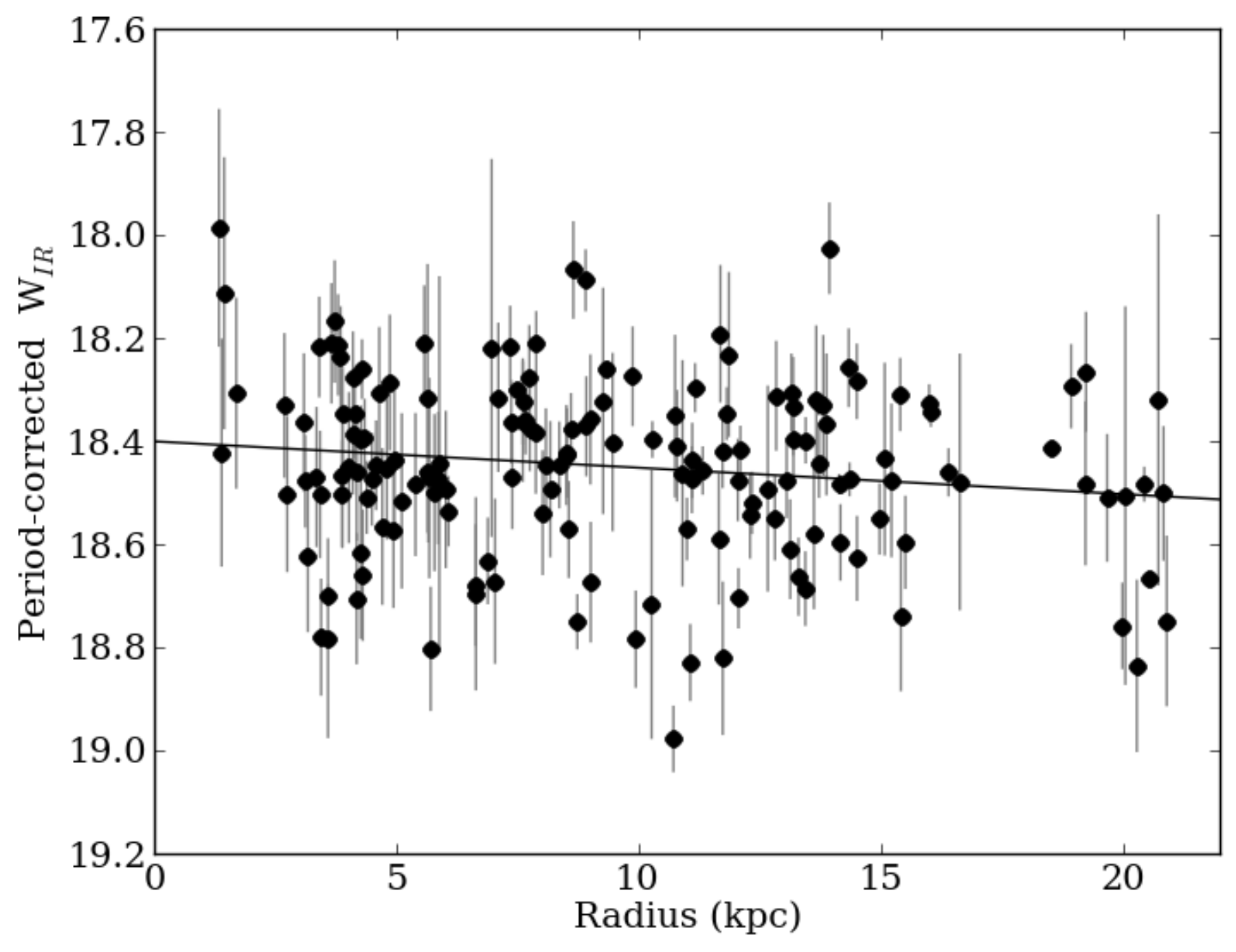}
\caption{The relationship between W$_{IR}$ magnitude and de-projected radius (using the distance modulus determined for the IR, $\mu_{IR}$=\NIRdist{}), where the magnitudes herein are adjusted to take into account the periods of the Cepheids, as described in the corresponding text, and include propagated errors. The linear fit has a slope of 0.005 $\pm$ 0.002. With a two-sided student t-test value of 1.89 and 159 degrees of freedom, the slope is significant beyond the 95\% level.}
\label{Wcorr}
\end{figure}

We also examine the radial magnitude trend in M31 by adjusting each Cepheid's Wesenheit magnitude based on its period, thereby correcting for the PL relations we determine in Sections \ref{IRPL} (this is comparable to examining the residual of the PL relation). The result of this exercise is plotted in Figure \ref{Wcorr}. We see that the Cepheids tend to be brighter in the inner region of the disk and fainter in the outer parts of M31. There is a correlation between magnitude and radial location with a two-sided student t-test at the 95\% significance level. We also apply Chauvenet's criterion to the data and find 7 outlying points with a significance level of 0.05; however, the slope is not significantly affected. The radial trend in magnitude is causing the change in distance modulus; however, the cause of the inwards brightening of Cepheids is unclear.

As previously discussed, we use the relation derived by Zaritsky et al. (1994) to calculate the 12+$\log_{10}$[O/H] values for each Cepheid as a proxy for the metallicity effects in M31. Zaritsky et al. (1994) found a radial metallicity gradient of --0.018$\pm$0.006 from their observations of HII regions in M31, which is quite shallow compared to many of the galaxies in that study. This is the gradient employed by this study and by Riess et al. (2009, 2011, 2012) in defining their distance scale. Sanders et al. (2012) find a metallicity gradient of --0.0195$\pm$0.005 in HII regions comparable to that found by Zaritsky; however, they do not find the trend echoed in their observations of planetary novae. Several studies have examined the globular clusters in M31 to search for a metallicity gradient. Barmby et al. (2000) looked at both spectroscopic and photometric metallicities of globular clusters, and while overall found no trend, observed a slight gradient of --0.023$\pm$0.01 when restricting their sample to clusters with spectroscopic metallicities. Huxor et al. (2011) use CMD metallicities but assert the lack of a metallicity gradient, suggesting that the gradient was driven primarily by metal-poor clusters. Freedman \& Madore (1990) use Cepheids in 3 different fields of M31 and, despite there being a slight difference in distance moduli between the inner and outer fields, conclude that their data is consistent with there being no dependence of the PL zeropoint on metallicity. However, Gould (1994) re-analyzed the same BVRI data used by Freedman \& Madore (1990) to show the M31 distance should be corrected for metallicity to avoid introducing additional errors or systematics. Lee et al. (2013) use 17 beat Cepheids to trace the metallicity in M31 and find a gradient of --0.008$\pm$0.004, shallower than that of Zaritsky et al. (1994) and Sanders et al. (2012), but similar to the PNe gradient from Kwitter et al. (2012) of --0.011 $\pm$0.004). While the metallicity gradient in M31 may not be particularly steep, there is a wealth of data supporting its existence and direction.

Previous empirical studies have suggested metallicity as the cause of the distance-radius relationship. However, there is disagreement between empirical studies and pulsation theory on how strong the metallicity effect is, as well as whether a higher metal content leads to a brighter or fainter Cepheid (Freedman \& Madore 1990; Kennicutt et al. 1998; Baraffe \& Alibert 2001; Macri et al. 2006; Caputo 2008; Bono et al. 2008, 2010; Romaniello et al. 2008, and references therein). Most theoretical approaches predict that metal-rich Cepheids should be fainter (Baraffe \& Alibert 2001; Caputo 2008). Conversely, most observational studies, including indirect and direct metallicity measurements, suggest that the more metal-rich Cepheids are brighter (Kovtyukh et al. 2005b; Macri et al. 2006; Bono et al. 2008), although Romaniello et al. (2008) presented spectroscopic evidence based on observations of Cepheid iron lines that disagrees with most empirical studies. An alternative viable explanation to metallicity effects could be crowding and blending, causing inner Cepheids to appear brighter (leading to shorter distances) compared to Cepheids in outer fields. 

Our results show a statistically significant distance-radius relationship in the IR bands, but without a significant metallicity component, suggesting that there may be another factor at least partially responsible for the difference. Majaess et al. (2011) argue that crowding is the cause of brighter Cepheids in the inner regions. In comparing ground-based and HST, Mochejska et al. (2000) and Chavez et al. (2012) both show that blending can be a significant effect, especially as seeing increases in the ground-based observations. Bono et al. (2008) that blending will be a stronger effect in the central region of a galaxy and thus lead to a decrease in distances.

Based on our results, crowding and blending could both be likely candidates for the cause of the brightening of inner Cepheids in the case of M31. We don't see a trend with magnitude and metallicity, but we do see a trend of magnitude (and distance) with radius. When we examine the photometric uncertainties from artificial star tests, we see the magnitude error and crowding parameter increase towards the center of M31. The average trend in uncertainty ($\sim$0.002 magnitudes/kpc) is sufficient to account for the observed gradient in magnitude and distance with radius. This suggests that a brightening bias due blending or crowding, rather than metallicity, could be the driving factor of the trend between radial distance and magnitude trends in M31.

\section{Discussion}\label{Discussion}

\begin{table*}
\centering
    \caption{M31 Distance Moduli}
    \begin{tabular}{cccccc}
    \hline
 \textbf{Distance modulus} &  \textbf{Filter(s)}	&	\textbf{Source}  	&	\textbf{Calibration}	&	\textbf{Anchor}	&	\textbf{Source} \\  
\hline
24.44$\pm{0.10}$	&	BVRI		&	Ground-based		&	Cepheids	&	LMC	&	Madore \& Freedman (1990) \\
24.44$\pm{0.12}$	& 	B, V 			&	Ground-based 	&	Eclipsing Binary	&	-	&	Ribas et al. (2005)	  \\  
24.32$\pm{0.12}$	& 	B, V			&	Ground-based	&	Cepheids	&	LMC	&	Vilardell et al. (2007)	  \\  
24.47$\pm{0.12}$	& 	I			& 	Ground-based	&	TRGB	&	-	&	Durrell et al. (2001)	  \\  
24.47$\pm{0.07}$	& 	I 			& 	Ground-based	&	TRGB	&	-	&	McConnachie et al. (2005)	  \\  
24.49$\pm{0.11}$	& 	R, I 			& 	Ground-based	&	Cepheids	&	LMC	&	Joshi et al. (2003)	  \\  
24.41$\pm{0.21}$	& 	R, I			&	Ground-based	&	Cepheids	&	LMC	&	Joshi et al. (2010)	  \\  
\hline
\VISdist{}				&	B, I			&		HST			&	Cepheids	&	This study		\\
24.5$\pm{0.1}$		& 	F606W, F814W &	HST			&	RR Lyrae	&	Carretta et al. (2000)	&	Brown et al. (2004)	  \\  
24.38$\pm{0.05}$	& 	V, I (F555W, F814W)	&	HST  	&	Cepheids	&	LMC/Key Project	&	Freedman et al. (2001)	  \\  
24.38 $\pm{0.064}$	& 	IR (F110W, F160W) &	HST		&	Cepheids	&	NGC 4258	&	Riess et al. (2012)	  \\  
\NIRdist{}				&	IR (F110W, F160W)	&	HST			&	Cepheids	&	This study		\\
24.54$\pm{0.08}$	& 	H (F160W)	&	HST  		&	Cepheids	&	LMC	&	Macri et al. (2001)	  \\  
        \hline
    \end{tabular}
   \label{distmoduli}
\end{table*}

\begin{figure}
\includegraphics[scale=0.45]{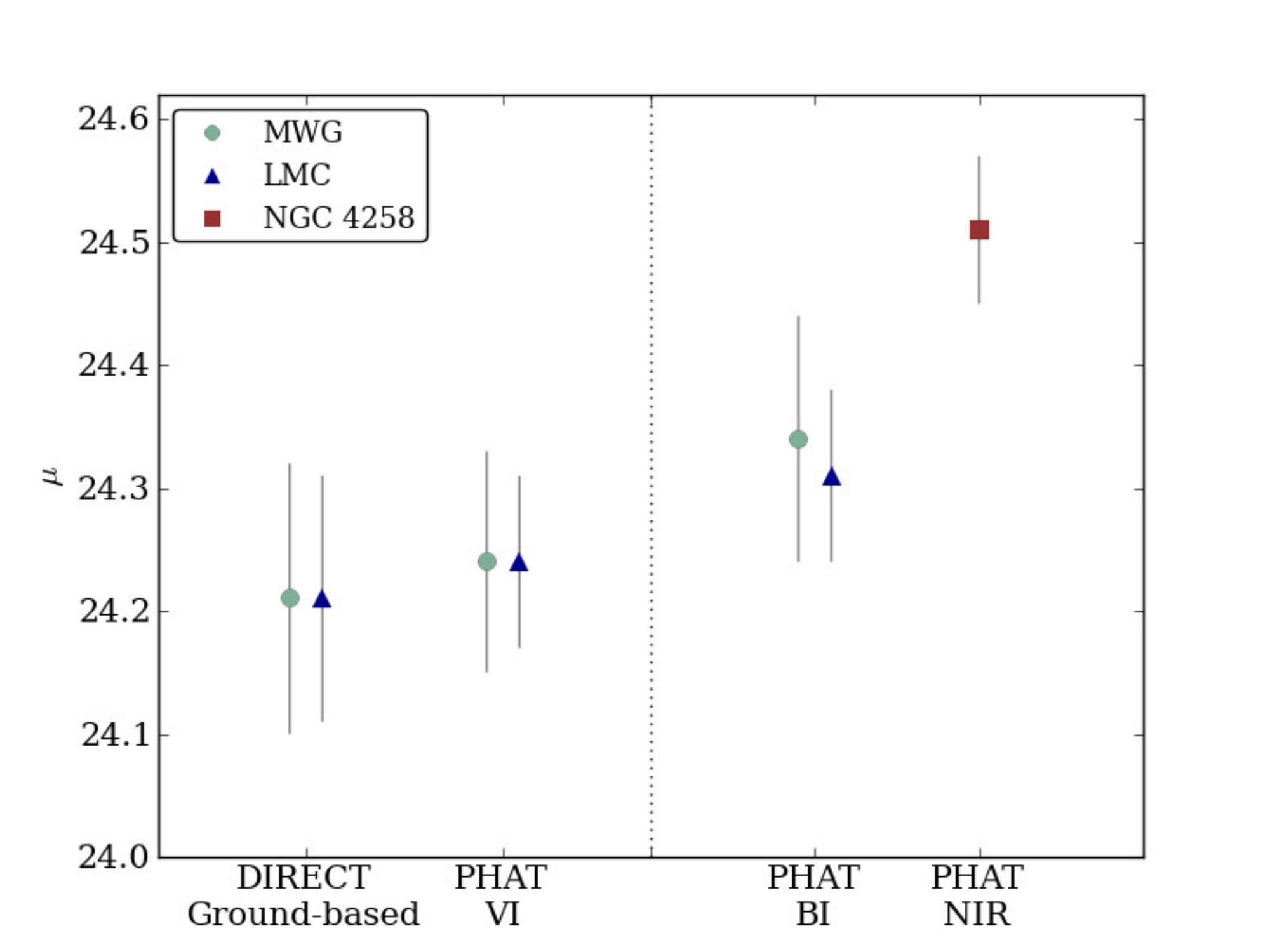}
\caption{The distance determinations from Table \ref{distmods} are plotted here, separated by filter. In the left half of the plot, the DIRECT distances and the PHAT distances using the DIRECT Cepheids are presented for both Galactic and LMC anchors. The the right half of the plot, the PHAT distances in the B and I filters are shown for both Galactic and LMC anchors, as well as the NIR distance determination.}
\label{distmods_mine}
\end{figure}

We present our distance estimates for the entire dataset (from Table \ref{Cepheidlist}) in Table \ref{distmods}. This dataset has the advantage of having the largest sample size of Cepheids in M31 in the visual bands from HST, as well as having a larger sample in the NIR bands than previous studies (Riess et al. 2012; Kodric et al. 2015).

Table \ref{distmods} and Figure \ref{slpint} show that there are significant biases in the slopes and intercepts of the PL relations of the ground-based DIRECT survey compared to the PHAT survey. The observed differences suggest that Cepheid observations obtained from ground-based surveys, while comprehensive in scope, are not ideal tools in the era of precision cosmology, and that Cepheid observations of more distant galaxies with HST may suffer from similar effects. Surprisingly, the biases do appear to cancel somewhat, leading to only very modest changes in the mean distance modulus.

In Table \ref{distmoduli}, we give recent values from other studies for the distance modulus of M31. For a full compilation and discussion of M31 distances, we refer the reader to de Grijs \& Bono (2014). The suggested M31 distance modulus from their study, incorporating various distance measurement techniques and statistically weighted errors, is 24.46$\pm$0.10 (assuming an LMC distance modulus of 18.50). Our distance modulus of \NIRdist{} determined from the NIR PL relations is in agreement with this range. 

In contrast to the NIR results, the visual Wesenheit (BI) distance modulus of \VISdist{} (an average of the MWG and LMC distances) falls outside of the 1-$\sigma$ range of de Grijs \& Bono (2014). The distance moduli determined from the W$_{BI}$ relations for the complete Cepheid sample and W$_{VI}$ relations for the DIRECT Cepheids with PHAT magnitudes are shorter than published distances, for both LMC and MWG anchors (see Table \ref{distmods}). They also disagree with our NIR distance determination of \NIRdist{}. Although our mean visual distance determination is within 1 sigma of several recent studies (Freedman et al. 2001, Ribas et al. 2005, Vilardell et al. 2007, Joshi et al. 2010, among others), it does fall on the short side of the range of recently published distances, as well as slightly outside the determination of de Grijs \& Bono (2014).

Unfortunately, there are no other similar published HST studies in comparable visual filters for Cepheids in M31 for a comparison. As examined by Mochejska et al. (2000), blending in the visual bands in ground-based observations may cause underestimates in Cepheid distances by 6 to  9\%. Although ground-based studies are more affected by these effects than HST, if further blending or crowding effects remain it could partially account for distances that remain shorter than expected. Williams et al. (2014) show that there is a brightening bias for the PHAT photometry in M31, especially for the inner region and fainter magnitudes. For a typical Cepheid, the infrared bands and F814W magnitudes can be biased up to 0.1 magnitudes (Williams et al. 2014). The F475W magnitudes are less affected, exhibiting a brightening of $\textless$0.05 magnitudes; colors also appear to have minimal bias at the level of a few hundredths of a magnitude. Magnitudes may be brightened up to 0.1 in the inner regions compared to $\sim$0.02 at 25 kpc from the center of M31. These biases reflect the crowding and blending problems present.

In a more direct comparison, we can compare our distance result in the infrared to similar NIR Cepheid studies in M31 with HST. Riess et al. (2012) obtain a distance modulus of 24.38$\pm$0.064. When we account for the increase of the distance to NGC 4258 by Humphreys et al. (2013) from 7.4 to 7.6 Mpc, this pushes the Riess et al. (2012) value to a 0.06 farther distance. This leads to a difference between the Riess et al. (2012) value and our NIR distance of 1-$\sigma$. The difference may be explained by differing global fits, where our slightly steeper slope and fainter intercept pushes towards a further distance.  Our distance estimate from the long period (greater than 10 days) Cepheids using the scale from Riess et al. (2009, 2011) is also consistent with the updated Galactic and LMC distance calibrations done by Bhardwaj et al. (2015) in the infrared. The distance determination by Kodric et al. (2015) gives a 0.068 closer distance to M31 relative to the Riess et al. (2012) value. This puts their distance approximately 1-$\sigma$ lower than Riess et al. 2012 and within 2-$\sigma$ of our distance estimate, when accounting for the difference in the updated maser distance. Kodric et al. (2015) obtain their Wesenheit magnitudes slightly differently, making a direct comparison incomplete. Additionally, they compare their PL fit to Riess et al. (2012) at log(P)=1.2, whereas we re-solve the system of M31 and NGC 4258 Cepheids. The final results of each study appear to be sensitive to either the included Cepheid sample, the distinct method for determining the distance, or both.

Derived from the complete sample of Cepheids, our NIR distance estimate of \NIRdist{} and visual distance of \VISdist{} disagree slightly beyond the 1-$\sigma$ level; it is unclear from where this disagreement stems. We do not find evidence for bias from random sampling and the optical bands face less crowding, as shown in Williams et al. 2014. It is also possible that blending from nearby companions has affected the visual filters more than the infrared filters, causing the visual distance estimate to be shorter. Previous studies suggest that the contamination due to blending may be greater in the visual than in the infrared, though more rigorous studies are needed to determine the effects and extent of blending on observations (Mochejska et al. 2000, Gieren et al. 2008).

We are inclined to trust the NIR distance estimate over the visual band estimate for several reasons. Though the visual bands are not in the same native photometric system as their calibrating relations from the Milky Way and LMC and the transformations to B, V, and I can be tricky. Although the NIR Wesenheit relation is also in a different photometric system than the original Riess et al. (2009) and Riess et al. (2011) papers, it is in the same system as Riess et al. (2012). Additionally, it is commonly thought that the longer wavelengths of the infrared reduce the effects of extinction, amplitude, and metallicity on Cepheid observations  (Madore \& Freedman 1991, Persson et al. 2004). Due to the absence of significant scatter in the Wesenheit BI relation, extinction effects are probably not severe. However, the remaining effects could introduce greater variations and possible biases in the visual bands than the infrared. Therefore, we suggest that our NIR distance estimate is a more reliable distance determination.


\section{Conclusions}\label{Conclusion}

We have analyzed Cepheid variables located in the PHAT data from the PAndromeda, DIRECT, and POMME (as available in Riess et al. 2012) ground-based surveys of M31. Through analyzing the visual and NIR magnitudes obtained from HST random phase observations, we make the following conclusions:

1. We use a sample of Cepheids to construct a Period-Luminosity relation using highly accurate HST magnitudes obtained by the PHAT survey. In particular, the visual Wesenheit relation from the complete sample (W$_{BI}$) shows less scatter over previous studies in the visual bands, leading to smaller random uncertainties in the distance modulus of M31.

2. The dispersions in the visual Wesenheit and the NIR Wesenheit relations are very similar. The dispersion in the NIR relations is comparable to that derived by Riess et al. (2012), but with a 2.5$\times$ larger Cepheid sample that further reduces the uncertainty in the M31 distance modulus. Although our dispersion is slightly larger than that of a similar study by Kodric et al. (2015), they use a different calculation of the Wesenheit magnitude.

3.  We obtain a value of the distance modulus for M31 of \VISdist{} from the visual filters and \NIRdist{} in the NIR using ACS and WFC3 photometry. These values are both consistent with recently published distance moduli. However, they disagree with each other by slightly more than 1-$\sigma$ due to the distance estimate at visual wavelengths being on the shorter end of published distances.

4. The PHAT survey provides highly accurate HST magnitudes of 175 Cepheids in M31. The superiority of the PHAT photometry from HST is clear in this dataset from the significantly smaller dispersions than that found using ground-based data, despite the fact that the Cepheid magnitudes are random phase. This is likely due to the enormous improvement in photometric quality and resolution as compared to ground-based surveys.

5. We find a statistically significant magnitude radial trend leading to a distance modulus-radius relationship, which does not appear to be explained by a metallicity correction but may be due to crowding or blending. Further work must be done to confirm or rule out metallicity as a cause of distance discrepancies between inner field and outer field Cepheids.


\section*{Acknowledgments}

The authors thank the entire PHAT team for their collaboration in this project, especially to M. Fouesneau, L. C. Johnson, and E. Skillman for their comments. We are grateful to A. Saha for numerous discussions during the development of the paper. We would also like to thank A. Riess for his input on some of the calculations herein. We thank L. Macri and A. Bonanos for their insightful comments on an earlier version of this paper. We also thank the anonymous referee who provided constructive comments. Support for this work was provided by NASA through GO-12055 from the Space Telescope Science Institute, which is operated by the Association of Universities for Research in Astronomy, Inc., under NASA contract NAS5-26555. RWK and AS thank the Indo-US Science and Technology Forum for support.





\begin{thebibliography}{10}

\bibitem[\protect\citeauthoryear{Allgood}{2006}]{b1} An, J. H., Evans, N. W., Hewett, P., et al. 2004, MNRAS, 351, 1071
\bibitem[\protect\citeauthoryear{Allgood}{2006}]{b1} Baade, W., \& Arp, H. 1964, ApJ, 139, 1027
\bibitem[\protect\citeauthoryear{Allgood}{2006}]{b1} Baraffe, I., \& Alibert, Y. 2001, A\&A, 371, 592
\bibitem[\protect\citeauthoryear{Allgood}{2006}]{b1} Barmby P., Huchra, J. P., Brodie, J. P., Forbes, D. A., Schroder, L. L., Grillmair, C. J., 2000, AJ, 119, 727
\bibitem[\protect\citeauthoryear{Allgood}{2006}]{b1} Benedict, G. F., McArthur, B. E., Feast, M. W. et al. 2007, AA, 133, 1810
\bibitem[\protect\citeauthoryear{Allgood}{2006}]{b1} Bhardwaj, A., Kanbur, S., Singh, H. P., et al. 2015, MNRAS, submitted.
\bibitem[\protect\citeauthoryear{Allgood}{2006}]{b1} Bonanos, A. Z., Stanek, K. Z., Sasselov, D. D., Mochejska, B. J., Macri, L. M., Kaluzny, J. 2003, AJ, 126, 175
\bibitem[\protect\citeauthoryear{Allgood}{2006}]{b1} Bono, G., Caputo, F., Fiorentino, G., Marconi, M., Musella, I. 2008, ApJ, 684, 102
\bibitem[\protect\citeauthoryear{Allgood}{2006}]{b1} Bono, G., Caputo, F., Marconi, M., Musella, I. 2010, ApJ, 715, 277
\bibitem[\protect\citeauthoryear{Allgood}{2006}]{b1} Bono, G., Marconi, M., Cassisi, S., Caputo, F., Gieren, W., Pietrzynski, G. 2005, ApJ, 621, 966
\bibitem[\protect\citeauthoryear{Allgood}{2006}]{b1} Brown, T. M., Ferguson, H. C., Smith, E., Kimble, R. A., Sweigart, A. V., Renzini, A., Rich, R. M. 2004, AJ, 127, 2738
\bibitem[\protect\citeauthoryear{Allgood}{2006}]{b1} Caputo, F. 2008, MmSAI, 79, 453
\bibitem[\protect\citeauthoryear{Allgood}{2006}]{b1} Caputo, F., Marconi, M., Musella, I. 2000, A\&A, 354, 610
\bibitem[\protect\citeauthoryear{Allgood}{2006}]{b1} Cardelli, J. A., Clayton, G. C., \& Mathis, J. S. 1989, ApJ, 345, 245
\bibitem[\protect\citeauthoryear{Allgood}{2006}]{b1} Carretta, E., Gratton, R. G., Clementini, G., \& Fusi Pecci, F. 2000, ApJ, 533, 215
\bibitem[\protect\citeauthoryear{Allgood}{2006}]{b1} Chavez J. M., Macri L. M., Pellerin A., 2012, AJ, 144, 113
\bibitem[\protect\citeauthoryear{Allgood}{2006}]{b1} Cotton, W. D., Condon, J. J., Arbizzani, E. 1999, ApJS, 125, 409
\bibitem[\protect\citeauthoryear{Allgood}{2006}]{b1} Dalcanton, J.; Williams, B.; Lang, D. et al. 2012, ApJS, 200, 18
\bibitem[\protect\citeauthoryear{Allgood}{2006}]{b1} de Grijs, R. \& Bono, G. 2014, AJ, 148, 17
\bibitem[\protect\citeauthoryear{Allgood}{2006}]{b1} Durrell, P. R., Harris, W. E., Pritchet, C. J. 2001, AJ, 121, 2557
\bibitem[\protect\citeauthoryear{Allgood}{2006}]{b1} Dolphin, A. 2000, PASP, 112, 1383
\bibitem[\protect\citeauthoryear{Allgood}{2006}]{b1} Fiorentino, G., Marconi, M., Musella, I., Caputo, F. 2007, A\&A, 476, 863
\bibitem[\protect\citeauthoryear{Allgood}{2006}]{b1} Fiorentino, G., Musella, I., Marconi, M., 2013, ArXiv e-prints, 1306.6276
\bibitem[\protect\citeauthoryear{Allgood}{2006}]{b1} Fliri, J., Riffeser, A., Seitz, S., Bender, R. 2006, A\&A, 445, 423
\bibitem[\protect\citeauthoryear{Allgood}{2006}]{b1} Fouqu\'e, P., Arriagada, P., Storm, J., et al. 2007, A\&A, 476, 73
\bibitem[\protect\citeauthoryear{Allgood}{2006}]{b1} Freedman, W. L., \& Madore, B. F. 1990, ApJ, 365, 186
\bibitem[\protect\citeauthoryear{Allgood}{2006}]{b1} Freedman, W. L., Madore, B. F., Gibson, B. K. et al. 2001, ApJ, 553, 47
\bibitem[\protect\citeauthoryear{Allgood}{2006}]{b1} Freedman, W. L., Madore, B. F. 2010, ARA\&A, 48, 673
\bibitem[\protect\citeauthoryear{Allgood}{2006}]{b1} Freedman, W. L., Madore, B. F., Scowcroft, V. 2012, ApJ, 758, 24
\bibitem[\protect\citeauthoryear{Allgood}{2006}]{b1} Gerke, J. R., Kochanek, C. S., Prieto, J. L., Stanek, K. Z.,  Macri, L. M. 2011, ApJ, 743, 176
\bibitem[\protect\citeauthoryear{Allgood}{2006}]{b1} Gieren W., Pietrzy\'nski G., Soszy\'nski I., Bresolin F., Kudritzki R.-P., Storm J., Minniti D., 2008, ApJ, 672, 266
\bibitem[\protect\citeauthoryear{Allgood}{2006}]{b1} Girardi, L., Bressan, A., Bertelli, G., Chiosi, C. 2000, A\&AS, 141, 371
\bibitem[\protect\citeauthoryear{Allgood}{2006}]{b1} Gould, A. 1994, ApJ, 426, 542
\bibitem[\protect\citeauthoryear{Allgood}{2006}]{b1} Haud, U. 1981, Ap\&SS, 76, 477
\bibitem[\protect\citeauthoryear{Allgood}{2006}]{b1} Herrnstein, J. R., et al. 1999, Nature, 400, 539
\bibitem[\protect\citeauthoryear{Allgood}{2006}]{b1} Hinshaw, G., Larson, D., Komatsu, E., et al. 2012, ArXiv e-prints, 1212.5226
\bibitem[\protect\citeauthoryear{Allgood}{2006}]{b1} Hodapp, K. W., Kaiser, N., Aussel, H., et al. 2004, AN, 325, 636
\bibitem[\protect\citeauthoryear{Allgood}{2006}]{b1} Humphreys, L., Reid, M., Moran, J., Greenhill, L., Argon, A., arXiv:1307.6031, 2013
\bibitem[\protect\citeauthoryear{Allgood}{2006}]{b1} Huxor, A. P., Ferguson, A. M. N., Tanvir, N. R., Irwin, M. J., Mackey, A. D et al. 2011, MNRAS, 414, 770
\bibitem[\protect\citeauthoryear{Allgood}{2006}]{b1} Joshi, Y. C., Pandey, A. K., Narasimha, D., Sagar, R., Giraud-Heraud, Y. 2003, A\&A, 402, 113
\bibitem[\protect\citeauthoryear{Allgood}{2006}]{b1} Joshi, Y. C., Narasimha, D., Pandey, A. K., \& Sagar, R. 2010, A\&A, 512, 66
\bibitem[\protect\citeauthoryear{Allgood}{2006}]{b1} Kaiser, N., Aussel, H., Burke, B. E., et al. 2002, Proc. SPIE, 4836, 154
\bibitem[\protect\citeauthoryear{Allgood}{2006}]{b1} Kaluzny, J., Mochejska, B. J., Stanek, K. Z., Krockenberger, M., Sasselov, D. D., Tonry, J. L., Mateo, M. 1999, AJ, 118, 346
\bibitem[\protect\citeauthoryear{Allgood}{2006}]{b1} Kaluzny, J., Stanek, K. Z., Krockenberger, M., Sasselov, D. D., Tonry, J. L., Mateo, M. 1998, AJ, 115, 1016
\bibitem[\protect\citeauthoryear{Allgood}{2006}]{b1} Kanbur, S. \& Ngeow, C., 2004, MNRAS, 350, 962
\bibitem[\protect\citeauthoryear{Allgood}{2006}]{b1} Kennicutt, R. C., Jr., Stetson, P. B., Saha, A., et al. 1998, ApJ, 498, 181
\bibitem[\protect\citeauthoryear{Allgood}{2006}]{b1} Kodric, M., Riffeser, A., Hopp, U., et al. 2013, AJ, 145, 106
\bibitem[\protect\citeauthoryear{Allgood}{2006}]{b1} Kodric, M., Riffeser, A., Seitz, S., et al. 2015, ApJ, 799, 144
\bibitem[\protect\citeauthoryear{Allgood}{2006}]{b1} Kovtyukh, V. V., Wallerstein, G., \& Andrievsky, S. M. 2005b, PASP, 117, 1173
\bibitem[\protect\citeauthoryear{Allgood}{2006}]{b1} Kwitter, K. B., Lehman, E. M. M., Balick, B., \& Henry, R. B. C. 2012, ApJ, 753, 12
\bibitem[\protect\citeauthoryear{Allgood}{2006}]{b1} Lee, C.-H., Kodric, M., Seitz, S., Riffeser, A., Koppenhoefer, J. et al. 2013, ApJ, 777, 35
\bibitem[\protect\citeauthoryear{Allgood}{2006}]{b1} Lee, C.-H., Riffeser, A., Koppenhoefer, J., et al. 2012, AJ, 143, 89
\bibitem[\protect\citeauthoryear{Allgood}{2006}]{b1} Leonard, D. C., Kanbur, S. M., Ngeow, C. C., Tanvir, N. R. 2003, ApJ, 594, 247
\bibitem[\protect\citeauthoryear{Allgood}{2006}]{b1} Macri, L. M. 2005, ArXiv e-prints, 0507648
\bibitem[\protect\citeauthoryear{Allgood}{2006}]{b1} Macri, L. M., Calzetti, D., Freedman, W. L., Gibson, B. K., Graham, J. A., et al., 2001, AJ, 549, 721 
\bibitem[\protect\citeauthoryear{Allgood}{2006}]{b1} Macri, L. M., Stanek, K. Z., Bersier, D., Greenhill, L. J., \& Reid, M. J. 2006, ApJ, 652, 1133
\bibitem[\protect\citeauthoryear{Allgood}{2006}]{b1} Madore, B. F. 1982, ApJ, 253, 575
\bibitem[\protect\citeauthoryear{Allgood}{2006}]{b1} Madore, B. F., \& Freedman, W. L. 1992, PASP, 104, 362
\bibitem[\protect\citeauthoryear{Allgood}{2006}]{b1} Madore, B. F., \& Freedman, W. L. 1991, PASP, 103, 933
\bibitem[\protect\citeauthoryear{Allgood}{2006}]{b1} Majaess, D., Turner, D.,  Gieren, W. 2011, ApJL, 741, L36
\bibitem[\protect\citeauthoryear{Allgood}{2006}]{b1} Marigo, P., Girardi, L., Bressan, A., Groenewegen, M. A. T., Silva, L., Granato, G. L. 2008, A\&A, 482, 883
\bibitem[\protect\citeauthoryear{Allgood}{2006}]{b1} McCommas, L. P., Yoachim, P., Williams, B. F., et al. 2009, AJ, 137, 4707
\bibitem[\protect\citeauthoryear{Allgood}{2006}]{b1} McConnachie, A. W., Irwin, M. J., Ferguson, A. M. N., Ibata, R. A., Lewis, G. F., Tanvir, N. 2005, MNRAS, 356, 979
\bibitem[\protect\citeauthoryear{Allgood}{2006}]{b1} Mochejska, B. J., Macri, L. M., Sasselov, D. D., Stanek, K. Z. et al. 2000, AJ, 120, 810
\bibitem[\protect\citeauthoryear{Allgood}{2006}]{b1} Moffett, T. J., \& Barnes, T. G., III 1986, ApJ, 304, 607
\bibitem[\protect\citeauthoryear{Allgood}{2006}]{b1} Monson, A. J., \& Pierce, M. J. 2011, ApJS, 193, 12
\bibitem[\protect\citeauthoryear{Allgood}{2006}]{b1} Ngeow C-C. 2012, ApJ, 747, 50
\bibitem[\protect\citeauthoryear{Allgood}{2006}]{b1} Ngeow, C.-C., \& Kanbur, S. M. 2005, MNRAS, 360, 1033
\bibitem[\protect\citeauthoryear{Allgood}{2006}]{b1} Ngeow, C-C, Kanbur, S.M., Nikolaev, S., Buonaccorsi, J., Cook, K., Welch, D. 2005, MNRAS 363, 831 
\bibitem[\protect\citeauthoryear{Allgood}{2006}]{b1} Opolski, A. 1983, Inf. Bull. Var. Stars, 2425, 1
\bibitem[\protect\citeauthoryear{Allgood}{2006}]{b1} Pejcha, O. \& Kochanek, C. 2012, ApJ, 748, 107
\bibitem[\protect\citeauthoryear{Allgood}{2006}]{b1} Persson, S. E., Madore, B. F., Krzeminski, W., et al. 2004, AJ, 128, 2239
\bibitem[\protect\citeauthoryear{Allgood}{2006}]{b1} Planck Collaboration et al. 2013, ArXiv e-prints, 1303.5076
\bibitem[\protect\citeauthoryear{Allgood}{2006}]{b1} Ribas, I., Jordi, C., Vilardell, F., et al. 2005, ApJ, 635, L37
\bibitem[\protect\citeauthoryear{Allgood}{2006}]{b1} Riess, A. G., Fliri, J., Valls-Gabaud, D. 2012, ApJ, 745, 156
\bibitem[\protect\citeauthoryear{Allgood}{2006}]{b1} Riess, A. G., Macri, L., Li, W. et al. 2009, ApJ, 183, 109
\bibitem[\protect\citeauthoryear{Allgood}{2006}]{b1} Riess, A. G., Macri, L., Casertano, S. et al. 2009b, ApJ, 699, 539
\bibitem[\protect\citeauthoryear{Allgood}{2006}]{b1} Riess, A. G., Macri, L., Casertano, S., et al. 2011, ApJ, 730, 119
\bibitem[\protect\citeauthoryear{Allgood}{2006}]{b1} Romaniello, M., Primas, F., Mottini, M., et al. 2008, A\&A, 488, 731
\bibitem[\protect\citeauthoryear{Allgood}{2006}]{b1} Saha, A., Shaw, R. A., Claver, J. A., Dolphin, A. E. 2011, PASP, 123, 481
\bibitem[\protect\citeauthoryear{Allgood}{2006}]{b1} Sanders, N. E., Caldwell, N., McDowell, J., \& Harding, P. 2012, ApJ, 758, 133
\bibitem[\protect\citeauthoryear{Allgood}{2006}]{b1} Simon, N. R. \& Lee, A., S. 1981, ApJ, 248, 291
\bibitem[\protect\citeauthoryear{Allgood}{2006}]{b1} Sandage, A., Tammann, G. A., Reindl, B. 2004, A\&A, 424, 43
\bibitem[\protect\citeauthoryear{Allgood}{2006}]{b1} Savage, B. D., \& Mathis, J. S. 1979, ARA\&A, 17, 73
\bibitem[\protect\citeauthoryear{Allgood}{2006}]{b1} Schlafly, E. F. \& Finkbeiner, D. P. 2011, ApJ, 737,103
\bibitem[\protect\citeauthoryear{Allgood}{2006}]{b1} Stanek, K. Z., Kaluzny, J., Krockenberger, M., Sasselov, D. D., Tonry, J. L., Mateo, M. 1998, AJ, 115, 1894
\bibitem[\protect\citeauthoryear{Allgood}{2006}]{b1} Stanek, K. Z., Kaluzny, J., Krockenberger, M., Sasselov, D. D., Tonry, J. L., Mateo, M. 1999, AJ, 117, 2810
\bibitem[\protect\citeauthoryear{Allgood}{2006}]{b1} Stanek, K. Z., \& Udalski, A. 1999, arXiv:astro-ph/9909346
\bibitem[\protect\citeauthoryear{Allgood}{2006}]{b1} Tammann, G. A. \& Reindl, B., Ap\&SS, 280, 165
\bibitem[\protect\citeauthoryear{Allgood}{2006}]{b1} Tammann, G. A., Reindl, B., Thim, F., Saha, A., Sandage, A. 2002, ASPC, 283, 258
\bibitem[\protect\citeauthoryear{Allgood}{2006}]{b1} Testa, V., Marconi, M., Musella, I., Ripepi, V., Dall'Ora, M., Ferraro, F. R., Mucciarelli, A., Mateo, M., C™tŽ, P. 2007, A\&A, 462, 599
\bibitem[\protect\citeauthoryear{Allgood}{2006}]{b1} Tonry, J., Onaka, P. 2009, in Proc. Advanced Maui Optical and Space Surveillance Technologies Conf., ed. S. Ryan (Kihei: Maui Economic Development Board), E40
\bibitem[\protect\citeauthoryear{Allgood}{2006}]{b1} Udalski, A., Szymanski, M., Kubiak, M., et al. 1999, Acta Astron., 49, 201
\bibitem[\protect\citeauthoryear{Allgood}{2006}]{b1} Vilardell, F., Jordi, C., Ribas, I. 2007, A\&A, 473, 847
\bibitem[\protect\citeauthoryear{Allgood}{2006}]{b1} Welch, D. L., Wieland, F., McAlary, C. W., McGonegal, R., Madore, B. F., McLaren, R. A., Neugebauer, G. 1984, ApJS, 54, 547
\bibitem[\protect\citeauthoryear{Allgood}{2006}]{b1} Williams, B., Dalcanton, J., Lang, D., et al. 2014, ApJ, 215, 9
\bibitem[\protect\citeauthoryear{Allgood}{2006}]{b1} Zaritsky, D., Kennicutt, R. C., Jr., Huchra, J. P. 1994, ApJ, 420, 87

\end{thebibliography}
\end{document}